\documentclass[aps,prl,superscriptaddress,groupedaddress,floatfix,longbibliography]{revtex4-1}
\usepackage[utf8]{inputenc}

\usepackage{graphicx}  
\usepackage{dcolumn}   
\usepackage{bm}        
\usepackage{amssymb}   
\usepackage{amsmath}   
\usepackage{amsthm}    
\usepackage{xcolor}    
\usepackage{tikz}      
\usepackage{ifthen}    
\usepackage{subfigure}
\usepackage{xspace}
\usepackage{lipsum}
\usepackage{gensymb}
\usepackage[english]{babel}
\usepackage{appendix}
\DeclareUnicodeCharacter{2212}{-}
\definecolor{violet}{RGB}{153,0,153}
\definecolor{teal}{RGB}{0,153,153}
\definecolor{darkorange}{RGB}{205, 98, 0}

\newcommand{\csrs}{(Ca$_x$Sr$_{1-x}$)$_3$Rh$_4$Sn$_{13}$}

\newcommand{\cro}{Cd$_2$Re$_2$O$_7$}
\newcommand{\srs}{Sr$_3$Rh$_4$Sn$_{13}$}
\newcommand{\xtec}{\textit{X-TEC}}

\begin{document} 
	
	\title{Harnessing Interpretable and Unsupervised Machine Learning to Address Big Data from Modern X-ray Diffraction} 
	
	\author{Jordan Venderley$^1$,
		Michael Matty$^1$,
		Krishnanand Mallaya$^1$,
		Matthew Krogstad$^2$,
		Jacob Ruff$^3$,\\
		Geoff Pleiss$^4$,
		Varsha Kishore$^4$,
		David Mandrus$^5$,
		Daniel Phelan$^2$,\\
		Lekh Poudel$^{6,7}$,
		Andrew Gordon Wilson$^8$,
		Kilian Weinberger$^4$,\\
		Puspa Upreti$^{2,9}$, M. R. Norman$^2$, Stephan Rosenkranz$^2$, Raymond Osborn$^2$,
		Eun-Ah Kim$^{1\ast}$\\
		\normalsize{$^{1}$Department of Physics, Cornell University}\\
		\normalsize{$^{2}$Materials Science Division, Argonne National Laboratory}\\
		\normalsize{$^{3}$Cornell High Energy Synchrotron Source, Cornell University}\\
		\normalsize{$^{4}$Department of Computer Science, Cornell University}\\
		\normalsize{$^{5}$Department of Materials Science and Engineering, University of Tennessee}\\
		\normalsize{$^{6}$Department of Materials Science and Engineering, University of Maryland}\\
		\normalsize{$^{7}$NIST Center for Neutron Research, National Institute of Standard and Technology}\\
		\normalsize{$^{8}$Courant Institute of Mathematical Sciences, New York University}\\
		\normalsize{$^{9}$Department of Physics, Northern Illinois University}\\
		\normalsize{$^\ast$To whom correspondence should be addressed; E-mail:  eun-ah.kim@cornell.edu.}
	}
	
	\date{\today}
	
	\baselineskip24pt
	
	\begin{abstract}
		The information content of crystalline materials becomes astronomical when collective electronic behavior and their fluctuations are taken into account. In the past decade, improvements in source brightness and detector technology at modern x-ray facilities have allowed a dramatically increased fraction of this information to be captured. Now, the primary challenge is to understand and discover scientific principles from big data sets when a comprehensive analysis is beyond human reach. 
		We report the development of a novel unsupervised machine learning approach, {\it XRD Temperature Clustering} (\xtec), that can automatically extract charge density wave (CDW) order parameters and detect intra-unit cell (IUC) ordering and its fluctuations from a series of high-volume X-ray diffraction (XRD) measurements taken at multiple temperatures.  We apply \xtec\ to XRD data on a quasi-skutterudite family of materials, (Ca$_x$Sr$_{1-x}$)$_3$Rh$_4$Sn$_{13}$, where a quantum critical point arising from charge order is observed as a function of Ca concentration.
		We further apply \xtec\ to XRD data on the pyrochlore metal, Cd$_2$Re$_2$O$_7$, to investigate its two much debated structural phase transitions and uncover the Goldstone mode accompanying them. 
		We demonstrate how unprecedented atomic scale knowledge can be gained when human researchers connect the \xtec\ results to physical principles. Specifically, we extract from the \xtec-revealed selection rule that the Cd and Re displacements are approximately equal in amplitude, but out of phase. This discovery reveals a previously unknown involvement of  $5d^2$ Re, supporting the idea of an electronic origin to the structural order. Our approach can radically transform XRD experiments by allowing in-operando data analysis and enabling researchers to refine experiments by discovering interesting regions of phase space on-the-fly. 
	\end{abstract}
	
	\maketitle 
	
	[1] From the early days of X-ray diffraction (XRD) experiments, they have been used to access atomic scale information in crystalline materials. The primary challenge has always been how to interpret the angle dependent scattering intensities of the resultant diffraction patterns (Fig 1(a)). Bragg and Bragg's initial insights into how to interpret such data \cite{Bragg} enabled the direct determination of crystal structures for the first time, and they were duly awarded a Nobel prize. Since the phase of the X-ray photon is lost in the measurement, the most common approach to interpreting XRD data is to employ forward modeling using the increasingly sophisticated tools of crystallography developed over the past century. These have been remarkably successful in determining the structure of highly crystalline materials, from simple inorganic solids to complex protein crystals. However, subtle structural changes can be difficult to determine when they only result in marginal changes in intensities without any change in peak locations \cite{egami:2012a}.
	Furthermore, thermal and quantum fluctuations captured in diffuse scattering away from the Bragg peaks are beyond the reach of conventional crystallographic analysis. 
	The information-rich diffuse scattering is typically weaker than Bragg scattering by several orders of magnitude and can be difficult to differentiate from background noise.
	
	[2] The massive data that modern facilities generate, spanning 3D reciprocal space volumes that include $\mathcal{O}(10^4)$ Brillouin zones (BZ) (Fig 1(a)), at rates of $\mathcal{O}(10^2)$ gigabytes per hour 
	should capture the systematics of such subtle atomic scale information. Yet the sheer quantity of data presents a major challenge. Overcoming this challenge is of paramount importance especially in searching for an unknown order parameter and its fluctuations. Specifically, two types of 
	orders and their fluctuations are targets of XRD (see the illustration for a one-dimensional system in Fig.~1(b-e)): those that change the size of the unit-cell, such as charge density waves (CDW), and those that involve intra-unit cell (IUC) distortions. XRD evidence of CDW order is the emergence of new superlattice peaks, which can be weak and fluctuating, often requiring a targeted search \cite{abbamonte:np2005a,forgan:nc2015a}. XRD evidence of IUC order are even subtler changes in structure factors of Bragg peaks~\cite{lawler:n2010a}, unless there are changes in extinction rules. However, the ubiquity of electronic nematic order \cite{Kivelson2010review,fernandes-review} has turned the study of electronically driven IUC order into an increasingly important scientific objective. Electronically driven IUC order and related `hidden order' phases typically have profound consequences for the electronic structure as revealed by various probes, yet are often accompanied by subtle structural distortions. Examples range from 3d oxides like cuprates, to 4d and 5d oxides like ruthenates and iridates, to 4f and 5f heavy fermion materials like URu$_2$Si$_2$.  These small distortions can challenge conventional crystallographic structural refinement that only tracks Bragg peaks and deduce the structural symmetry by fitting all the atomic positions in a forward model.
	As an example of proposed CDW order, the quasi-skutterudite family, (Ca$_x$Sr$_{1-x}$)$_3$X$_4$Sn$_{13}$ where X is a transition metal ion like Co, Rh or Ir, exhibits marginal Fermi liquid behavior.  Much like in cuprates and heavy fermion materials such as YbRh$_2$Si$_2$, this order can be suppressed to very low temperatures, leading to a linear in temperature resistivity over a large range in temperature.
	In the pyrochlore, Cd$_2$Re$_2$O$_7$, a very subtle structural distortion is associated with large changes in the specific heat and susceptibility.  This led Liang Fu \cite{Fu:2015cl} to propose the presence of spin nematic order, and some evidence for this was provided by subsequent non-linear optics measurements \cite{Harter2017}.
	Moreover, the inversion breaking structural order itself is novel, whose candidate description by an $E_u$ tensor could support pseudo-Goldstone fluctuations between its two components, $I4_{1}22$ and $I\bar{4}m2$ (see Fig.~1(f)) \cite{kendziora:prl2005a}. 
	Interestingly, both of these examples exhibit superconductivity at low temperatures, leading to the question of how superconductivity is related to these orders.
	
	[3] To extract atomic scale information encoded in massive XRD data volumes,
	much needed is a versatile, interpretable, and scalable approach that can reveal order parameters and fluctuations associated with CDW orders and IUC orders: the vision behind \xtec. For the analysis of complex experimental data, dimension reduction and machine learning techniques are increasingly employed~\cite{Zhang2019,Bohrdt2019,ghosh:sa2020a,Torlai2019,ronhovde:sr2012a,ziatdinov:cm2020a,Geddes2019Structural,Wright2017Computer}, with an emphasis on supervised learning using hypothesis-driven synthetic data \cite{Zhang2019,Bohrdt2019,ghosh:sa2020a}. 
	To date, most applications of unsupervised techniques to materials data have been limited to exploration of compositional phase diagrams of alloys~\cite{long2009rapid,stanev2018unsupervised,Chen2021Machine}. 
	However, an interpretable and unsupervised approach aiming at discovering interaction driven emergent phenomena in quantum materials such as order parameters and fluctuations can greatly benefit scientific progress.
	For versatility, we opted for an unsupervised approach guided by a fundamental principle of statistical mechanics: the balance between the energy ($E$) and entropy ($S$) resting on the temperature ($T$).   
	A change in the collective state of a system occurs in the direction of minimizing the Helmholtz free energy $F$~\cite{beale2011statistical}:
	\begin{equation}
		F=E-TS.
		\label{eq:F}
	\end{equation}
	When the temperature $T$ is lowered below a certain threshold, the entropy $S$ gives way to the ordered state dominated by the system Hamiltonian. 
	Hence the temperature ($T$) evolution of the XRD intensity for reciprocal space point $\vec{q}$, $I(\vec{q}, T)$, must be qualitatively different if the given reciprocal space point $\vec{q}$ reflects order parameters or their fluctuations.
	Tracking the temperature evolution of thousands of Brillouin zones to identify systematic trends and correlations in any comprehensive manner is impossible to achieve manually without selection bias (see Fig.~2(a)). \xtec\ embodies the principle of Eq.~\ref{eq:F} by clustering the `temperature series' associated with a given $\vec{q}$, $I(\vec{q},T)$, according to qualitative features in the temperature dependence, as in high-dimensional clustering approaches that learn qualitative differences in the voice trains for speaker verification~\cite{reynolds2000speaker} (see Fig.~2(b)). \xtec\ achieves interpretability and scalability by
	using a simple Gaussian mixture model (GMM)~\cite{Murphy2013} at its core (see SM section II) and incorporates correlation among nearby $\vec{q}$ points and within and across BZs using label smoothing (see SM section II-C) similar to how signals from different cameras can be correlated for computer vision \cite{you:apa2020a} (Fig.~2(c)). 
	
	[4]
	The first step in the \xtec\ pipeline is to preprocess the raw set of temperature series obtained for each reciprocal space point $\{\vec{q}=(q_x,q_y,q_z)\}$ in $\sim 10^9$ grid points in 3D reciprocal space grid over 10-30 temperatures. The challenges in working with the raw comprehensive data are in the volume and the dynamic range of the intensity scale (see Fig.~2(a)). Our preprocessing scheme (SM section II-B) reduces the number of $\vec{q}$-space points to be canvassed from the full grid to a selection of points $\{\vec{q}_i\}$ through a systematic thresholding. The trajectories that passed the thresholding, $\{I(\vec{q}_i, T_j), j=1,\cdots,d^T\}$, are then rescaled so that we can compare trajectories at different intensities scales, focusing on their temperature dependence rather than the absolute scale. We now cluster the resulting collection of preprocessed temperature trajectories,  ${\bf \tilde{I}}({\vec{q}_i})\equiv\{\tilde{I}(\vec{q}_i, T_j); j=1,\cdots,d^T\}$ for each $\vec{q}_i$ 
	to discover qualitatively distinct types of temperature dependences in the data. 
	There are two modes for clustering depending on whether the label smoothing is on or off: \xtec ~smoothen (\xtec s) and \xtec~detailed (\xtec d).  
	For \xtec s, we adopt the label smoothing approach that effectively correlates signals from different cameras for comupter vision \cite{you:apa2020a} to correct the 
	independence assumption and enforce local smoothness across the cluster assignments of points with similar momenta within and across Brillouin zones. The algorithm first constructs a nearest neighbor graph in momentum space, connecting reciprocal space points that share similar momenta. For each point, the neighbors are weighted by their distance in momentum space and the weights normalized. Label smoothing averages the cluster assignments of a point with its (weighted) neighbors. We incorporate this smoothing step between the E- and M- step of the GMM. The human researcher makes the choice between \xtec s and \xtec d
	and trades between a simpler output and a simpler (i.e., more scalable) algorithm. \xtec s is best suited for detecting order parameters while \xtec d can reveal the nature of fluctuations in high-resolution data. Using \xtec s and \xtec d in tandem can reveal systematic correlations between order parameters captured by peak centers and fluctuations captured by diffuse scattering in an unprecedented manner.  
	
	[5] Outside of the label smoothing, 
	\xtec\ uses standard GMM on the temperature series, $\{{\bf \tilde I}(\vec{q}_i)\}$, treated as a point in the $d^T$-dimensional space. Once the human researcher sets the number of clusters $K$,  \xtec\ attempts to model  
	each point in the data set $\{{\bf \tilde I}(\vec{q}_i)\}$  to be independently and identically drawn from a weighted sum of $K$ distinct multivariate normal distributions. The hyper-parameters to be learned are the mixing weights $\pi_k$, $d^T$-dimensional means ${\bf m}_k$, and $d^T\times d^T$-dimensional covariances  ${\bf s}_k$, $(\pi,{\bf m},{\bf s})\equiv\{(\pi_k,{\bf m}_k,{\bf s}_k); k=1,\cdots,K\}$. The associated model log-likelihood is
	\begin{equation}
		\log p\Big( \{{\bf \tilde I}(\vec{q}_i)\} | \pi,{\bf m},{\bf s}\Big) = \sum\limits_{\vec{q}_i} \log \left[ \sum\limits_{k = 1}^K \pi_{k} \mathcal{N}\Big( {\bf \tilde I}(\vec{q}_i)| {\bf m}_k,{\bf s}_k\Big) \right].
		\label{eq:likelihood}
	\end{equation}
	Here,
	$\mathcal{N}\Big( {\bf \tilde I}(\vec{q}_i)| {\bf m}_k,{\bf s}_k\Big)$ is the probability density for the $k^{th}$ multivariate Gaussian with mean ${\bf m}_k$ and covariance ${\bf s}_k$ evaluated at ${\bf \tilde I}(\vec{q}_i)$, i.e.,  
	\begin{equation}
		\mathcal{N}\Big({\bf \tilde I}(\vec{q}_i) | {\bf m}_k, {\bf s}_k\Big) \equiv \frac{1}{(2 \pi)^{d_T/2}} \frac{1}{\sqrt{\det {\bf s}_k}} e^{-\frac{1}{2}\Big[{\bf \tilde I}(\vec{q}_i - {\bf m}_k)^{\dagger} {\bf s}_k^{-1} ({\bf \tilde I}(\vec{q}_i) - \mu_k)\Big]}.
		\label{eq:gaussian}
	\end{equation}
	The probability, $w_i^k$, that the temperature series labeled by $\vec{q}_i$ belongs to the $k^{th}$ cluster is
	\begin{equation}
		w_i^k  = \frac{\pi_k \mathcal{N}\Big({\bf \tilde I}(\vec{q}_i) | {\bf m}_k, {\bf s}_k\Big)}{\sum\limits_k \pi_k \mathcal{N}\Big({\bf \tilde I}(\vec{q}_i) | {\bf m}_k, {\bf s}_k\Big)},
		\label{eq:expectation}
	\end{equation}
	according to Bayes' theorem (see SM section IIc).
	\xtec\ learns the hyper-parameters  $(\pi, {\bf m}, {\bf s})$ 
	using a stepwise expectation maximization (EM) algorithm \cite{Liang2009}.
	Much like mean-field theory familiar to physicists, the EM algorithm iteratively searches for the saddle point of the lower bound of the log-likelihood
	\begin{equation}
		\tilde{\ell}\Big(\{w^k_i, \pi_k, {\bf m}_k, {\bf s}_k\}\Big) = \sum\limits_{i, k} w_i^k \log\Bigg[ \frac{\pi_k \mathcal{N}\Big({\bf \tilde I}(\vec{q}_i) | {\bf m}_k, {\bf s}_k\Big)}{w_i^k} \Bigg] + \lambda (1 - \sum\limits_k \pi_k),
	\end{equation}
	where $\lambda$ is a Lagrange multiplier. The cluster assignment of a given reciprocal space point $\vec{q}_i$ is then determined by the converged value of the clustering expectation 
	$\mathop{\mathrm{arg\,max}}_k \{ w^k_i \}$. 
	
	[6] We first employ \xtec s to target a putative CDW quantum critical point and illustrate \xtec s in action. 
	Electrical resistivity and heat capacity experiments on the quasi-skutterudite family \csrs\ indicate a quantum critical point at a composition of $x = 0.9$ under ambient pressure (see Fig.~2(g)) \cite{Goh2012}, driven by the volume change of replacing the larger Sr ion by the smaller Ca one, with superconductivity emerging at low temperatures.
	Given that this is associated with a linear in temperature resistivity as in the cuprates,
	the question has arisen about the nature of this order, the quantum critical fluctuations associated with it, and their connection to the superconductivity.  Although CDW order was proposed \cite{Goh2012}, this has never been proven, so we use \xtec s to investigate this.
	The x-ray measurements on \csrs\ were taken on Sector 6-ID-D at the Advanced Photon Source using a monochromatic x-ray energy of 87 keV. Images are collected on a fast area detector (Pilatus 2M CdTe) at a frame rate of 10 Hz while the sample is continuously rotated through 360\degree\ at a speed of 1\degree\ per second (Fig. 1a). These rotation scans are repeated twice to fill in gaps between the detector chips, so a single measurement represents an uncompressed data volume of over 100 GB collected in 20 minutes. This allows comprehensive measurements of the temperature dependence of a material in much less than a day. Using a cryostream, we are able to vary the temperature from 30 K to 300 K. The rotation scans sweep through a large volume of reciprocal space (Fig. 1a); when the data are transformed into reciprocal space coordinates, the 3D arrays are typically reduced in size by an order of magnitude. More details of both the measurement and data reduction workflow are given in Ref.~\cite{Krogstad:2019tc}, see also SM I. 
	In the past, we would have analyzed such data by selecting a few superlattice peaks, with the assumption that they are representative of the whole, and fitting their temperature dependence. This may be justified in many cases, but in doing so, we would be ignoring over 99\% of the data, limiting the statistical precision available from such comprehensive data sets and potentially missing other components of the order parameter.
	Here we apply \xtec s to around 200 GB of XRD data on four compounds, $(x=0, 0.1, 0.6, 0.65)$ and map out the phase diagram as a function of temperature and doping with no prior knowledge regarding the order parameter given to \xtec s.

	[7] \xtec s extracts order parameter clusters from the entire 200 GB of XRD data for the four compounds within minutes. In Figs.~2(d) we present cluster means and variances of the two-cluster ($K=2$) clustering results for undoped Sr$_3$Rh$_4$Sn$_{13}$.
	The temperature dependence of the learned means of the yellow cluster and the blue cluster makes it evident that the yellow cluster represents the order parameter and the temperature at which it crashes down is the critical temperature: $T_c\approx 130$ K.
	The clustering results can be interpreted by locating the cluster assignments in reciprocal space, as shown in Fig.~2(e). The location of the yellow cluster identifies the ordering wavevector to be $q_{CDW}$ = (0.5,0.5,0) and symmetry equivalents with respect to the cubic Bragg peaks, without any prior knowledge.  
	Label smoothing keeps the clustering output to be smoothly connected in the vicinity of each peak, simplifying interpretation. Plotting the CDW order parameters extracted at each doping, we can track the evolution of the critical temperature $T_c$ as a function of chemical pressure (Fig.~2(f)) and obtain the full quantum phase diagram. The doping-dependent $T_c$ obtained using \xtec\ allows us to map out the quantum phase diagram associated with the CDW ordering, much as neutron scattering has been used to obtain quantum phase diagrams associated with spin order. Earlier studies of this family of compounds identified quantum critical behaviour using 
	thermodynamic, transport, and phonon measurements \cite{Goh:2015cl,cheung:pr2018a}, but this is the first to be determined directly from the CDW order parameter and shows the efficiency with which X-TEC analysis can extract structural phase diagrams. The suppression of the CDW order upon doping and the emergence of superconductivity from a strange metal with linear $T$ resistivity is reminiscent of the cuprates \cite{Keimer}. 
	
	[8] We now employ \xtec s and \xtec d in tandem to study hidden IUC order and order parameter fluctuations in  the 
	pyrochlore metal Cd$_2$Re$_2$O$_7$ \cite{Jin.2001, Hanawa.2001, Sakai.2001} (see Fig.~3(a)),
	where the nature of its two $E_u$-symmetry structural transitions have recently regained interest \cite{Hiroi2018} after the discovery of a purported $T_{2u}$ electronic order from second harmonic generation (SHG) \cite{Harter2017}.
	Cd$_2$Re$_2$O$_7$ goes through a second-order transition at $T_{s1}=200$ K with a large thermodynamic signature in the specific heat (Fig.~3(b)) from the  cubic pyrochlore $Fd\bar{3}m$  structure (phase I) to a structure that breaks inversion symmetry (phase II). Most studies conclude that the space group of the phase II is $I\bar{4}m2$ \cite{Hiroi2018}, but this is now questioned in light of the SHG data \cite{Harter2017,di-matteo:pr2017a,Norman:2020hn}, which also reveal the surprising fact that the $E_u$ structural order (which it also sees) does not have the expected temperature dependence of a primary order parameter (unlike the $T_{2u}$ signal, which does).
	At lower temperature, a first-order transition at $T_{s2}=113$ K (phase III) has been observed, and is proposed to arise from the other component of $E_u$ which is the $I 4_122$ space group \cite{Hiroi2018}.  Again, this is controversial, in that earlier SHG data \cite{petersen:np2006a} do not see the expected rotation of the signal that should accompany such a phase transition.
	Moreover, recent Raman data \cite{kapcia:prr2020a} see line splittings consistent with a lowering to orthorhombic symmetry below about $80$ K \cite{kapcia:prr2020a} which was speculated to be due to an $F222$ space group.
	A combination of small atomic displacements with crystallographic twinning \cite{Castellan:2002kz} has made it challenging to determine the true structure of these low symmetry states using traditional crystallographic approaches \cite{Yamaura:2002fk,yamaura:pr2017a}.
	Still, previous results for phase II are consistent with the above picture, where $I\bar{4}m2$ and $I 4_122$ are the two components of the $E_u$ order parameter, a rank-2 tensor. The degeneracy between these two states is lifted at sixth order in Landau theory \cite{Ivan:2003dl}, resulting in a quasi-Goldstone mode encoding fluctuations between the two phases \cite{goldstone, meier:pr2020a} (see Fig.~1(f)). Raman scattering \cite{kendziora:prl2005a} sees a strong central peak that appears to be the Goldstone mode, along with a higher frequency mode which appears to be the Higgs mode, though this has been recently questioned based on pump-probe measurements \cite{harter:prl2018a}. The uniqueness of this situation is that although pseudo-Goldstone modes have been seen in other materials, notably ferroelectrics, they typically exist at much higher frequencies \cite{meier:pr2020a}.  The fact that this is not the case for Cd$_2$Re$_2$O$_7$ indicates that the anisotropy in the Landau free energy is anomalously small.  Confirmation of such low frequency fluctuations has been beyond the reach of XRD, as has been the relation of the $E_u$ structural order to the proposed $T_{2u}$ `hidden order' indicated by the SHG data.
	
	[9] We performed x-ray scattering measurements over 
	a wide temperature range ($30$ K $<T<300$ K) on a single crystal of Cd$_2$Re$_2$O$_7$, which our measurements show is untwinned, at least in phase II. This may be due to the small volume (400x200x50 $\mu$m$^3$) required for our synchrotron measurements. We first performed scans using an x-ray energy of 87 keV, which contained scattering spanning nearly 15,000 Brillouin zones, in order to search for previously undetected peaks and determine the systematic ($HKL$) dependence of the Bragg peak intensities at each temperature (see SM section III-A). To better understand the order parameter fluctuations, we then reduced the energy to 60 keV to improve the $\vec{Q}$-resolution and increased the number of temperatures, particularly near the phase transitions. We comprehensively analyzed the resulting data sets with a combined volume of nearly 8 TB using \xtec s and \xtec d.

	[10] We illustrate the sharp characteristics of the order parameter and its fluctuations by focusing on the cubic-forbidden peaks in Figures 3 and 4 (see SM III-B for the clustering results that selects cubic-forbidden peaks as the order parameter of phase II).
	Fig.~3(c) shows the $K=2$ clustering means of \xtec s and $K=3$ clustering means of \xtec d  on the cubic-forbidden peaks over the temperature range of [30 K,150 K] \footnote{For each \xtec\ clustering we increase $K$ until there is no gain in information.}.  Both outcomes presented big surprises. First, the \xtec s outcome separated the cubic forbidden peaks that behave like the order parameter of phase II into two subgroups: one  that quickly flattens in phase II to abruptly rise in phase III  (yellow) and the other that continues to rise in phase II to abruptly drop in phase III (green). Second, 
	\xtec d clustering separates out 
	the diffuse regions associated with each of the subgroups of cubic-forbidden peaks to define their own clusters with temperature dependencies that are qualitatively different (red and blue in Fig.~3(c)) and distinct from the temperature dependencies of the peak centers. 
	The reciprocal space distribution of the clusters reveals precise selection rules and tight correlation between the order parameter tracked in \xtec s and the fluctuations revealed in \xtec d. Due to the orders of magnitude differences in intensity scales, \xtec s is dominated by the peak centers. \xtec d separated out the peak centers from the halos of diffuse  regions. Combining the two results, we present the \xtec s outcome through the color of the peak centers detected in \xtec d. The $(HKL)$ assignments of the two subgroups  in \xtec s, and their associated diffuse halos in \xtec d (Fig.~3(d)) reveal strict selection rules. Yellow peaks (with red halos) are of the form $(4n_1, 4n_2, 4n_3+2)$, while green peaks (with blue halos) have $(4n_1+2, 4n_2, 4n_3)$ or $(4n_1, 4n_2+2, 4n_3)$, in the cubic indices of phase I.
	The mean intensity trajectories of red and blue clusters in Fig.~3(c) indicate that the red halo sustains intensity throughout phase II to only dive down at $T_{s2}=113$ K while the blue halo picks up intensity at around $T_{s2}$ to abruptly die out at around $90$ K. The temperature evolution of representative line cuts shown in Fig.~3(e-f) confirms these observations in the raw data.

	[11] The systematics in the temperature dependencies of different cubic-forbidden peaks and their diffuse halos revealed using the two modes of \xtec\ on the entire 8 TB of data present an unprecedented opportunity to extract atomic scale clues regarding the hidden order. First, we can extract an order parameter critical exponent associated with the structural transition that is reflecting the entire data set from the \xtec s mean trajectories. Fig.~4(a) shows the temperature dependence of the two peak averaged clusters (yellow and green) of cubic-forbidden peaks and their fits, in which we treat the displacements as order parameters with a common exponent $\beta$ (see SM III-D). 
	Both clusters fit to the common 
	exponent of $\beta\approx 0.25$ close to $T_{s1}$. This is close to the value expected for a 2D-XY system \cite{Bramwell:1993ig}. This is a surprise in that the $E_u$ signal observed by SHG scales linearly in $T_{s1}-T$ which is $4\beta$ instead of the expected $2\beta$ indicated by theory \cite{di-matteo:pr2017a}, whereas it is the $T_{2u}$ signal that scales like $2\beta$.  Second, we can convert the selection rule revealed by \xtec~into atomic distortions. 
	The selection rule shows that the two clusters correspond to two distinct classes of structure factor, whose values only depend on the distortions of the Cd and Re sublattices: the yellow cluster consists of peaks that are dominated by $z$-axis displacements $(\delta z_{\rm{Cd}},\delta z_{\rm{Re}})$, and those in the green cluster by in-plane displacements, along $x$ or $y$ depending on the Wyckoff position, $(\delta x_{\rm{Cd}},\delta x_{\rm{Re}})$ (SM III-C) (see Fig.~4(b)). The flat temperature dependence of the yellow cluster below 180 K results from out-of-phase distortions of the Cd and Re sublattices. The refined values of $(\delta z_{\rm{Cd}}$ and $\delta z_{\rm{Re}})$ are approximately equal and opposite (see Fig.~4(b)). This is another surprising result.  Previous refinements \cite{Weller} indicate that the Re displacements are small, and this is consistent with a density functional theory study \cite{kapcia:prr2020a}. 
	Small Re displacements are expected if the $5d$ electrons in Re play a passive role in the structural transition as the Re are in an almost ideally bonded octahedral environment,  compared to Cd which is underbonded because of its two short Cd-O and six long Cd-O bonds.  Therefore, a large displacement of Re implies that this is a consequence of the $5d^2$ configuration of Re being unstable to spin nematic order that should lead to valence bond ordering (different Re-Re bonds, as illustrated in Fig.~1(f)) in a given Re tetrahedron as proposed in other pyrochlores \cite{Tchernyshyov}.
	Third, the connection between the two diffuse halo clusters (red and blue) and the selection rule for the peak centers draws us to the unusual and distinct temperature dependence of the diffuse regions (see Fig.~4(c)). Strong critical scattering at $T_{s1}$ is clear in both clusters, but the diffuse contribution is much stronger in the red halo throughout phase II. The role between the two halos reverses at $T_{s2}$. We attribute the fluctuations reflected in the sustained intensity of the red halo to the Goldstone mode manifest through strong   $z$-axis fluctuations. 
	
	[12] To investigate this further, we turn to a description of the various modes (see SM III-F for more details of the calculations).  Above $T_{s1}$ one has a soft mode whose energy should go to zero at $T_{s1}$.  Below this, the soft mode splits into a Higgs mode (fluctuations in the amplitude of the $E_u$ order) and a Goldstone mode (fluctuations in the phase, that is fluctuations between $I\bar{4}m2$ and $I4_122$).  The latter would be at zero energy if there were no anisotropy.  In Landau theory, the first anisotropy term appears at sixth order and the next one at eighth order in the free energy.  These two must be of opposite sign in order to have a second transition at $T_{s2}$ \cite{Ivan:2003dl}.  Their difference changes sign at $T_{s2}$. The net result is that one has a Goldstone mode that starts at zero energy at $T_{s1}$, rises slightly with lowering $T$, then dips down again at $T_{s2}$, and then rises again below this.  This can be appreciated by the intensities associated with the various modes (see Fig.~4(d)), noting that the Goldstone mode's coupling to the x-rays is quadratic in the $E_u$ order parameter \cite{Fleury} reflecting the fact that it does not exist above $T_{s1}$ (the analog of the soft mode below $T_{s1}$ is the Higgs mode).  From the calculated intensities, one sees that the Goldstone mode completely dominates outside of the critical region near $T_{s1}$.  The calculated behavior is remarkably similar to the XRD data (Fig.~4(c)), with a pronounced cusp at $T_{s2}$.  This is strong indication that the diffuse scattering is indeed due to structural fluctuations associated with the Goldstone mode.
	
	[13] In summary, we developed \xtec, an unsupervised and interpretable ML algorithm for voluminous XRD data that is guided by the fundamental role temperature plays in emergent phenomena.  
	By analyzing the entire data set over many BZs and making use of temperature evolutions, \xtec\ can pick up subtle features representing both order parameters and fluctuations from higher intensity backgrounds. The two modes, \xtec s and \xtec d, allow for discovery of systematics in order parameters and its fluctuations despite orders of magnitude differences in intensities.
	The algorithm is fast with $O(10)$ minutes of run time for the tasks presented here. Using \xtec, we discovered that the superconductor family (Ca$_x$Sr$_{1-x}$)$_3$Rh$_4$Sn$_{13}$ exhibits CDW order and we mapped out its phase diagram. In Cd$_2$Re$_2$O$_7$, we conclusively identified the primary order parameter of the $T_{s1}=200$ K transition. We further revealed 
	the nature of the intra-unit-cell atomic distortions in a way that has eluded crystallographic analysis until now. 
	Finally, we revealed XRD evidence of a structural Goldstone mode for the first time. 
	The unprecedented degree of microscopic information we have been able to unearth from the XRD is fitting for such comprehensive data but would have been impossible by manual inspection. Instead of determining critical exponents by fitting a handful of peaks, \xtec\ provides a means of including the entire data volume by clustering peak intensities from thousands of Brillouin zones to produce an analysis that is both robust and rapid in future studies of such phase diagrams.  Given the general structure of \xtec, we anticipate it to be broadly applicable to other fields beyond XRD.  
	
	\nocite{ng_lecturenotes}
	\bibliography{X-ray-ML_sources}
	
	\section*{Acknowledgments}
	We acknowledge the assistance of Anshul Kogar in the TiSe$_2$ measurements. We thank 
	Jeffrey Lynn and Johnpierre Paglione for assistance in preparing the \csrs samples.
	Initial development of \xtec\ (EAK, AW, KW, GP) was supported by NSF HDR-DIRSE award number OAC-1934714 
	and testing on TiSe$_2$ data was supported by U.S. Department of Energy, Office of Basic Energy Sciences, Division of Materials Science and Engineering under Award DE-SC0018946 (JV).  The experiments on \csrs\ and \cro\ (MK, SR, RO, PU, DP), and the subsequent machine learning analysis and theoretical interpretations of the results (EAK, VK, JV, MN, KM), were supported by the US DOE, Office of Science, Office of Basic Energy Sciences, Division of Material Sciences and Engineering.
	MM acknowledges support by the National Science Foundation (Platform for the Accelerated Realization, Analysis, and Discovery of Interface Materials (PARADIM)) under Cooperative Agreement No. DMR-1539918 and the Cornell Center for Materials Research with funding from the NSF MRSEC program (DMR-1719875). This research used resources of the Advanced Photon Source, a US DOE Office of Science User Facility operated for the DOE Office of Science by Argonne National Laboratory under Contract No. DE-AC02-06CH11357. Research conducted at CHESS is supported by the National Science Foundation via Awards DMR-1332208 and DMR-1829070.
	
	\clearpage
	\section*{Figure Captions}
	\includegraphics[width=\linewidth]{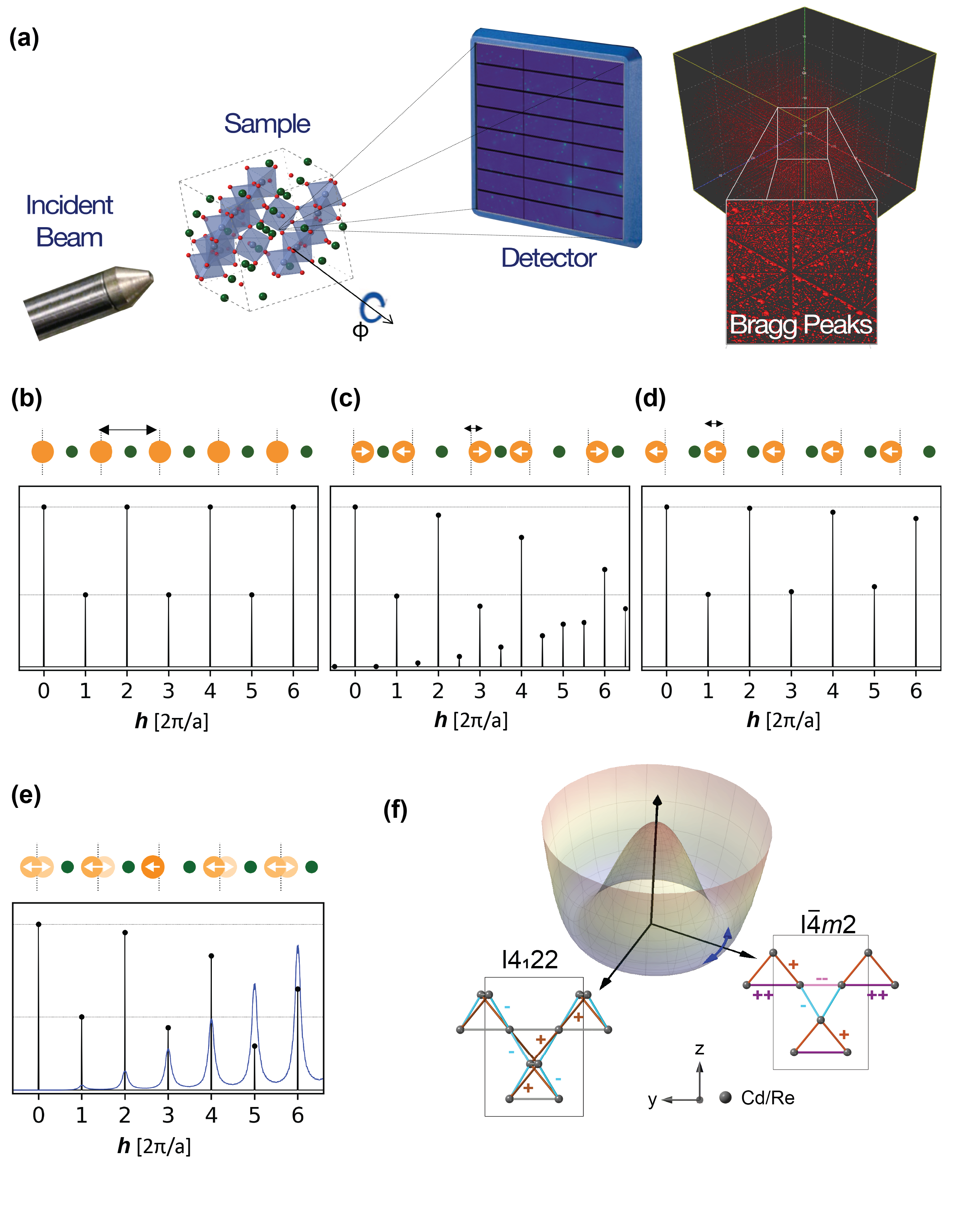}
	\noindent{\bf Fig. 1} 
	{\bf (a)} Schematic geometry of the x-ray scattering measurements. A monochromatic x-ray beam is incident on the sample, which rotates about the orthogonal $\phi$ axis while images are captured on a fast area detector. The reciprocal space map shows the $\vec{q}$-coverage of a single plane in the 3D volume after capturing images over a full $360^\circ$ sample rotation. A three-dimensional volume of reciprocal space covered by the x-ray scattering is shown on the right. Each red dot is a single Bragg peak. With an x-ray energy of 87 keV, a volume of over 10,000 \AA$^{-3}$ is measured, containing over ten thousand Brillouin zones if the unit cell dimension is 10 \AA.
	{\bf (b-e)} Real space positions of atoms (top) and the corresponding scattering intensities (bottom) calculated from simulated one-dimensional crystals with a unit cell containing two atoms, illustrating (b) a high symmetry phase, with (c) distortions due to CDW order, (d) IUC order and (e) short-range IUC order. In (b), the high symmetry phase produces peaks at integer $\vec{q}$. In (c), displacements of the orange atoms by $\pm\delta$ double the size of the unit-cell producing additional super-lattice peaks at half-integer $\vec{q}$ as well as changes in the other peak intensities. In (d), IUC distortions of the orange atoms by $-\delta$ change the peak intensities without producing additional super-lattice peaks. In (e), every orange atom is displaced by $\pm\delta$, with a 70\% probability of nearest neighbors having the same displacement. This finite correlation length has a small impact on the total scattering (black), but produces broad diffuse scattering (blue, x70000 scale compared to total scattering).   
	{\bf (f)} Bond patterns on the pyrochlore lattice associated with an $E_u$ distortion as inferred in Cd$_2$Re$_2$O$_7$.  The two space groups refer to the two different components of $E_u$ with each bond color denoting a different bond length. The amount of distortion of each bond from the average bond (grey) is indicated by $++,-,$ etc. along with the respective bond color.\\
	\includegraphics[width=\linewidth]{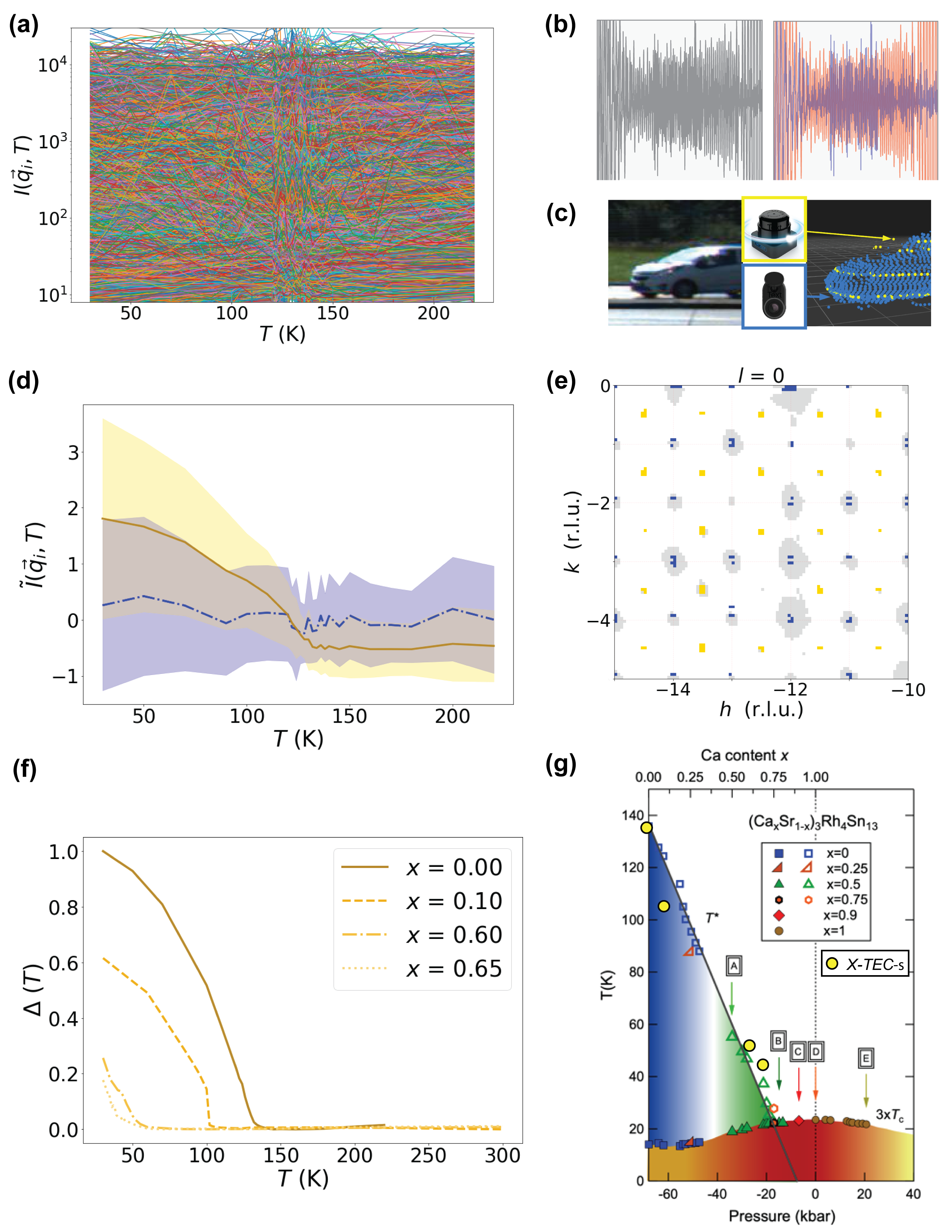}
	{\bf Fig 2:} \xtec~ with label smoothing (\xtec s) on \csrs. {\bf(a):} Example of raw intensity trajectories for \srs. The plot shows the collection  of individual
	raw temperature series $I(\vec{q}_i,T)$ for each point $\vec{q}_i$ in the data set spanning the reciprocal space $(h,k,l=0)$ where $h, k\in [-15,15]$ reciprocal lattice units (r.l.u.). {\bf (b)} Sound waveform of two people simultaneously talking (left) can be separated through clustering represented by different colors (right). {\bf (c)} Performing depth estimation for self driving cars, aggregating multiple sensor information with label smoothing. Depth estimation from LIDAR (yellow) are highly accurate but sparse, while depth estimation from cameras (blue) are dense but noisy. Label smoothing synthesizes the two sources, aligning the noisy camera observations to match LIDAR observations~\cite{you:apa2020a}.
	{\bf (d)} Two-cluster results of XRD data from \srs~ with the clustering assignments color-coded as yellow and blue.  Each raw intensity trajectories of (a) are re-scaled [$\tilde{I}(\vec{q}_i,T)$] by dividing their individual mean over temperature  and subtracting one, before clustering. The lines represent cluster means and the shaded region shows one standard deviation, interpolated between $24$ temperature points of measurement.  {\bf(e)} The corresponding yellow/blue cluster assignments of the $\vec{q}_i$ pixels that passed the thresholding. The image is zoomed to a section of the $(h,k,0)$ plane. The low intensity background (white) and the  $\tilde{I}(\vec{q}_i,T)$ with low temperature variance (grey) are eliminated by thresholding (see SM. II-B). {\bf (f)} The cluster means of the CDW clusters are interpolated and plotted to reveal  order parameter $\Delta(T)$ like behavior for four samples at different
	values of Ca doping $x$. $\Delta(T)$ is estimated from  the cluster means by subtracting the minimum from each cluster mean and appropriate normalization. 
	{\bf (g)}
	The critical temperatures estimated from $\Delta(T)$ (yellow  filled circles) overlaid onto the known phase diagram from \cite{Goh:2015cl} based on phase boundaries from thermodynamic measurements and transport.\\
	
	\includegraphics[width=\linewidth]{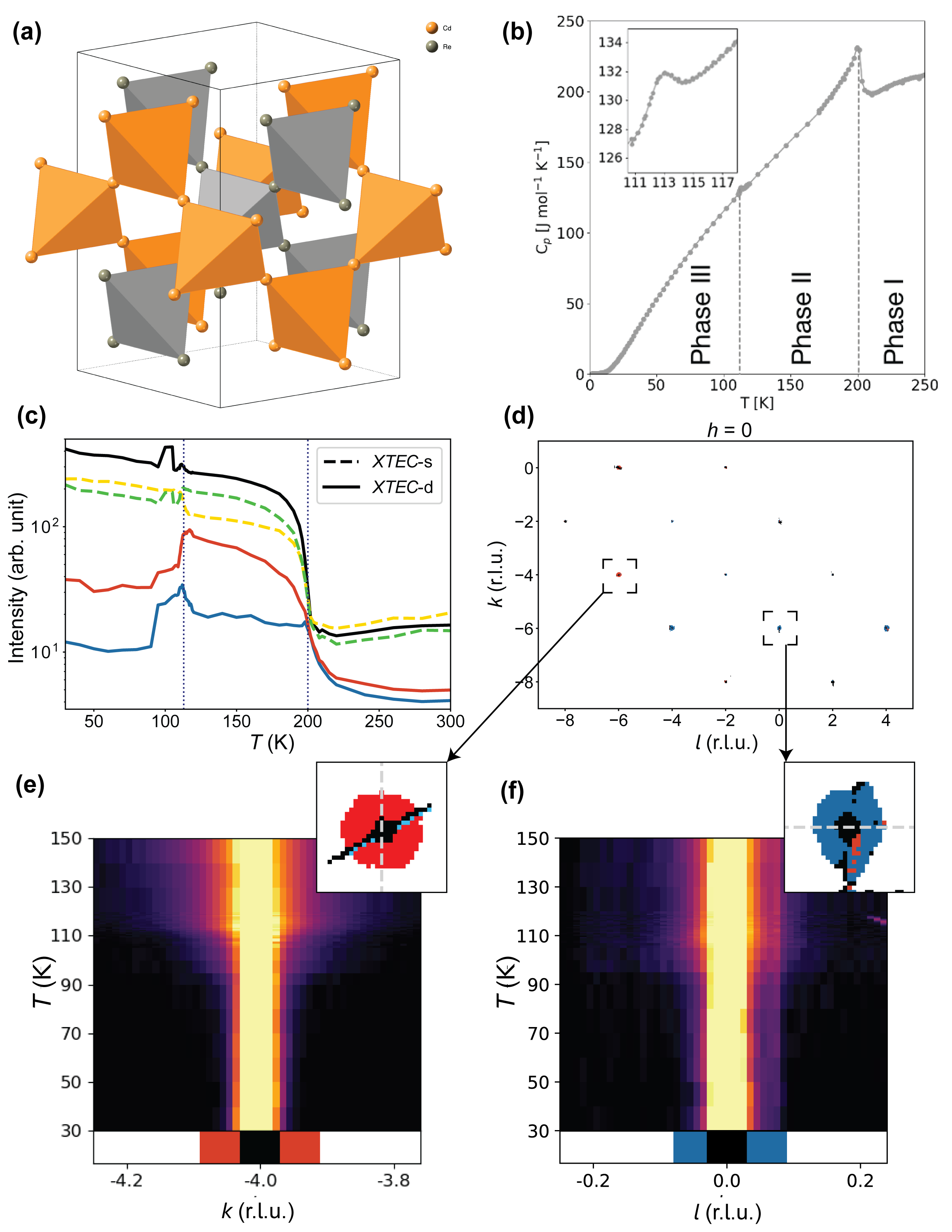}
	\noindent{\bf Fig. 3} \xtec\ analysis of \cro. {\bf (a)} 
	Crystal structure of \cro\ showing only Cd and Re, in the high temperature cubic phase.
	{\bf (b)} 
	Temperature dependence of the specific heat of \cro, showing the second-order phase transition at $T_{s1}$=200 K and the first-order phase transition at $T_{s2}$=113 K (see SM III-A). Three temperature ranges are marked as phase I ($T>T_{s1}=200K$), phase II ($T_{s2}=113K<T<T_{s1}$), and phase III ($T<T_{s2}$).
	{\bf (c)} 
	\xtec\ results on the cubic forbidden Bragg peaks from high resolution XRD data, showing temperature dependence of the mean intensity of each cluster (the cluster assignments are obtained from  30 K $\le T\le$ 150 K data, see SM III-C for details). The solid lines show three-cluster $(K=3)$ \xtec d trajectories, color coded as black, red and blue. The dashed line shows two-cluster $(K=2)$ \xtec s (peak averaged) trajectories, colored yellow and green. The temperatures of the two structural phase transitions are shown as dotted lines. {\bf (d)} The \xtec d cluster-color assignments of the thresholded pixels in a section of the  $h=0$ plane, where $k$ and $l$ are in reciprocal lattice units (r.l.u.). The pixels are assigned black, red and blue colors as in (c). The regions in the vicinity of two Bragg peaks at $0\overline{46}$ (left) and $0\overline{6}0$ (right) are magnified to show that the peak centers in both belong to the black cluster while halos form two distinct clusters  (red and blue respectively) separated from their peak centers. {\bf (e-f)} The raw intensity plotted for $0\overline{46}$ (left) and $0\overline{6}0$ (right) along a line cut (the grey dashed line shown in the respective zoom-ins) confirm the temperature dependence of the red  and blue halo intensities represented by the cluster means in (c). Specifically, 
	the $0\overline{46}$ peak has enhanced diffuse scattering above $T_{s2} \approx 113$ K, consistent with the temperature dependence of the red cluster mean. The $0\overline{6}0$ peak shows an anomaly near $T_{s2}$ and a suppressed diffuse scattering above, consistent with the temperature dependence of the blue cluster mean. 
	
	\includegraphics[width=\linewidth]{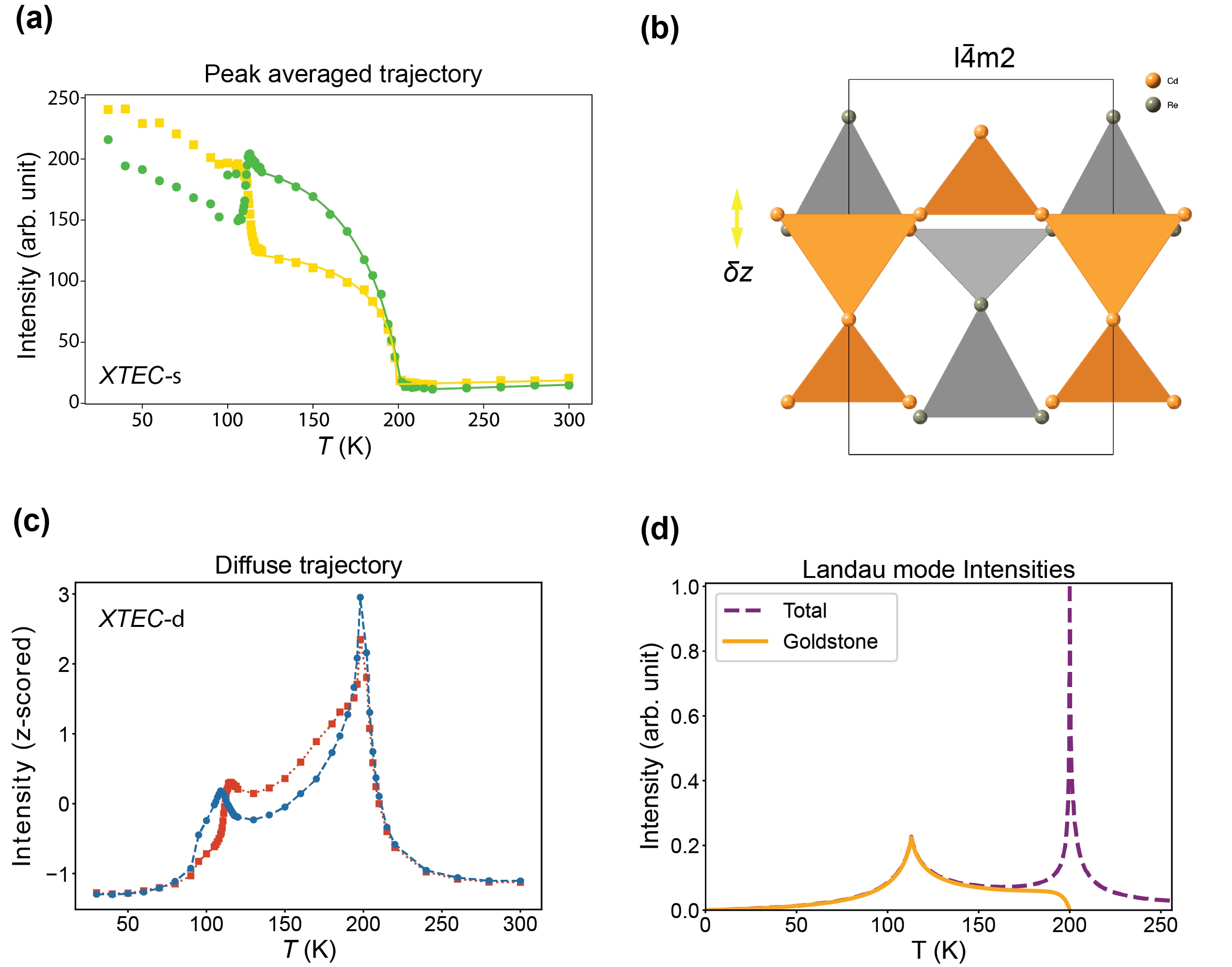}
	\noindent{\bf Fig. 4}
	Order parameters and their fluctuations inferred from \xtec\ outcomes. {\bf (a)} The filled symbols are the two-cluster mean intensity trajectories of peak averaged data [yellow and green trajectories from Fig. 3(c)], and solid lines are fits to these cluster means based on the model assuming $\delta x$ displacements (yellow) and $\delta z$ displacements (green) 
	of cations to vary as $(T-T_c)^{\beta}$, with a common order parameter exponent of $\beta=0.25$ as discussed in SM Section III-D.
	{\bf (b)} Schematic diagram of the relative $z$-axis displacements of cation sublattices for the $Cd$ (orange) and $Re$ (gray)  with respect to the cubic phase, inferred from the fit in (a). The \xtec-discovered selection rule and the fit establish the approximately equal magnitude but out-of-phase displacements $\delta z_{Cd}$  and $\delta z_{Re}$. {\bf (c)} The characteristic temperature dependence of the diffuse clusters are revealed by the z-scored intensities (for each intensity, subtract their mean over $T$ and then divide their standard deviation in $T$). The red and blue trajectories correspond to the respective cluster average of the z-scored intensities. Lines are guides for the eyes. The symbols square (circle) in (a) and (c) indicate that red (blue) diffuse clusters are associated with yellow (green) Bragg peaks. {\bf (d)} The calculated Landau mode intensities as a function of $T$ (see SM III-F).  Outside of the critical region near $T_{s1}$ (200 K), the intensity is dominated by the Goldstone mode intensity.  Note the resemblance of the calculated intensity to the diffuse trajectory in (c). 
	
\end{document}


\title{Harnessing Interpretable and Unsupervised Machine Learning to Address Big Data from Modern X-ray Diffraction}
	\author{Jordan Venderley$^1$,
		Michael Matty$^1$,
		Krishnanand Mallayya$^1$,
		Matthew Krogstad$^2$,
		Jacob Ruff$^3$,\\
		Geoff Pleiss$^4$,
		Varsha Kishore$^4$,
		David Mandrus$^5$,
		Daniel Phelan$^2$,\\
		Lekh Poudel$^{6,7}$,
		Andrew Gordon Wilson$^8$,
		Kilian Weinberger$^4$,\\
		Puspa Upreti$^{2,9}$, M. R. Norman$^2$, Stephan Rosenkranz$^2$, Raymond Osborn$^2$,
		Eun-Ah Kim$^{1\ast}$\\
		\normalsize{$^{1}$Department of Physics, Cornell University}\\
		\normalsize{$^{2}$Materials Science Division, Argonne National Laboratory}\\
		\normalsize{$^{3}$Cornell High Energy Synchrotron Source, Cornell University}\\
		\normalsize{$^{4}$Department of Computer Science, Cornell University}\\
		\normalsize{$^{5}$Department of Materials Science and Engineering, University of Tennessee}\\
		\normalsize{$^{6}$Department of Materials Science and Engineering, University of Maryland}\\
		\normalsize{$^{7}$NIST Center for Neutron Research, National Institute of Standard and Technology}\\
		\normalsize{$^{8}$Courant Institute of Mathematical Sciences, New York University}\\
		\normalsize{$^{9}$Department of Physics, Northern Illinois University}\\
		\normalsize{$^\ast$To whom correspondence should be addressed; E-mail:  eun-ah.kim@cornell.edu.}
	}

	\date{\today} 
	
	\maketitle
	
	\section*{Supplementary Information}
	
	\section{X-ray Measurements}
	A schematic of the x-ray measurement scattering geometry is shown in Fig. 1 of the main article. Three-dimensional volumes of diffuse X-ray scattering were collected at Advanced Photon Source (APS) and CHESS. The APS data were measured on sector 6-ID-D using an incident energy of 87.1 keV and a detector distance of 638 mm, except for the high-resolution measurements on \cro, which used an incident energy of 60.0 keV and a distance of 1406 mm. The raw images were collected on a Dectris Pilatus 2M with a 1-mm-thick CdTe sensor layer. The data were collected over a temperature range of 30 K to 300 K, with samples cooled by flowing He gas below 105 K and N$_2$ gas above 105 K.  The CHESS data on TiSe$_2$ were measured on beamline A2 using an incident beam energy of 59 keV and a Dectris Pilatus 6M detector with a 1-mm-thick Si sensor layer. The data were collected over a temperature range of 90 K to 300 K, with samples cooled by flowing N$_2$ gas. During the measurements, the samples were continuously rotated about an axis perpendicular to the beam at 1\degree s$^{-1}$ over 360\degree, with images read out every 0.1 s. Three sets of rotation images were collected for each sample at each temperature to fill in gaps between the detector chips. The resulting images were stacked into a three-dimensional array, oriented using an automated peak search algorithm and transformed in reciprocal space coordinates using the software package CCTW (Crystal Coordinate Transformation Workflow), allowing S(\textbf{Q}) to be determined over a range of $\sim\pm$15~\AA$^{-1}$ in all directions ($\sim\pm$6~\AA$^{-1}$ for the high-resolution measurement on Cd$_2$Re$_2$O$_7$). Further details are given in ref. \citenum{Krogstad:2019tc}.
	
	\section{\xtec: XRD Temperature    Clustering}\label{label:ML_xray}
	
	\subsection{\xtec d analysis of TiSe$_2$ CDW ordering}
\begin{figure}[H]
		\centering
		\includegraphics[width=\linewidth]{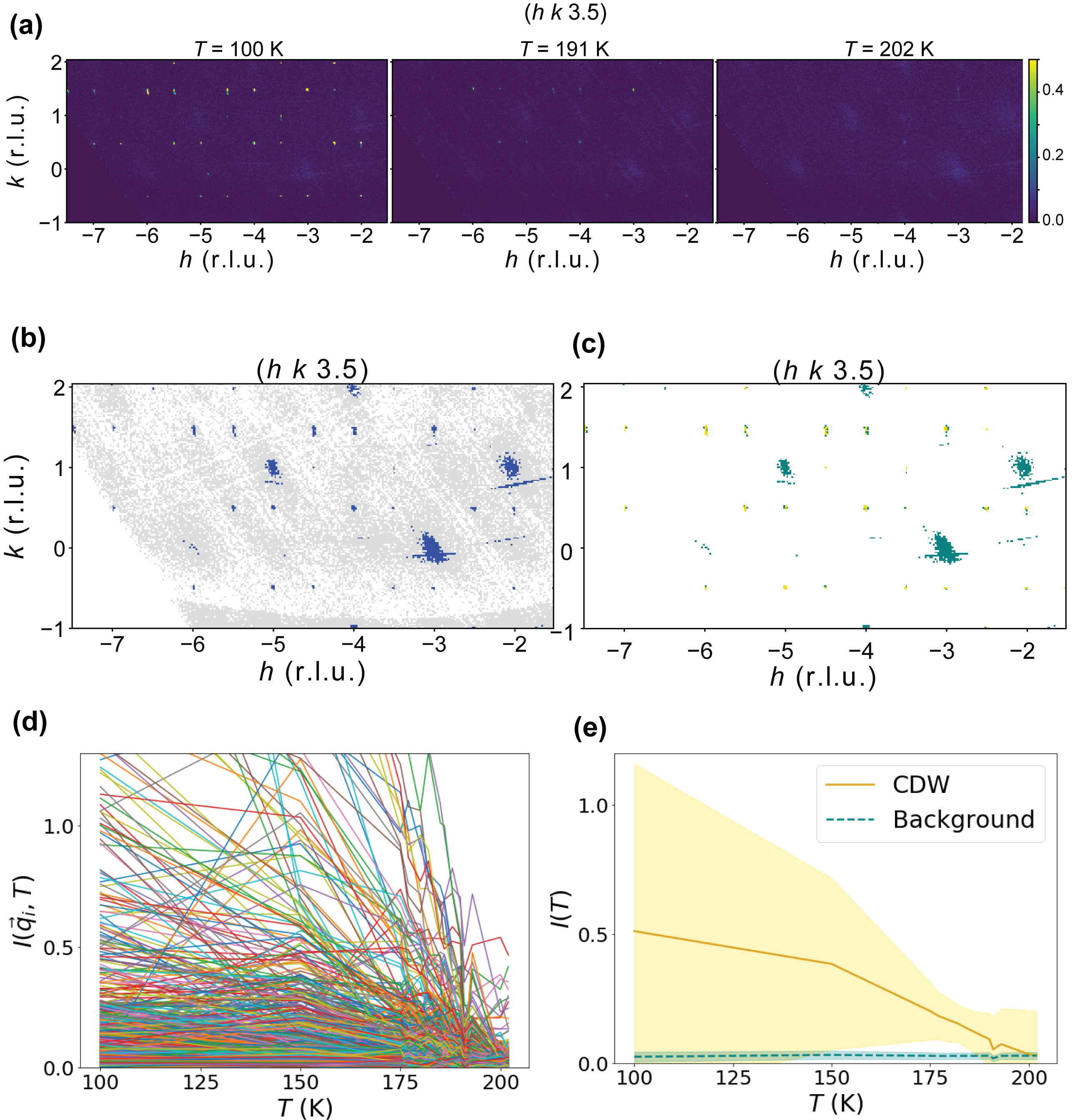}
		\caption{\xtec d analysis of TiSe$_2$. {\bf (a)} Two-dimensional slices of intensity of 1T-TiSe$_2$ on the $l\!=\!3.5$ plane at three temperatures. This plane contains super-lattice peaks at $T<T_c=200K$ (left) that disappears with the melting of the CDW order (right). $(h,k,l)$ are in reciprocal lattice units (r.l.u.), and the color-map over saturates intensity (arb. unit.) $> 0.5$. 
			{\bf (b)} Thresholding described 
			in SM.~\ref{preprocessing} removes the grey clusters in the reciprocal space of the plane shown in (a). Only the blue clusters belonging to a set $\{\vec{q}_i\}$ are tracked using \xtec.
			{\bf (c)} \xtec d two-cluster $(K=2)$ results assign the colors yellow and teal to the blue pixels of (b). The locations of yellow pixels identify with the CDW peaks, while teal pixels identify with the background scattering. {\bf (d)} Raw non-Bragg intensity trajectories over $d^T=14$ temperature values, $\{T_1 = 100K, \cdots, T_{14} = 202K\}$, of all $\vec{q}_i$-points 
			in a $3\times3\times3$ BZ's used for clustering.
			{\bf (e)} Temperature trajectories of the two clusters from the GMM analysis of data in (d). Lines denote cluster means $\mathbf{m}$ and shading represents covariance $\mathbf{s}$ (one standard deviation) for the non-trival CDW cluster (yellow) and the background cluster (teal), interpolated between the $d^T=14$ temperatures measured.  \\
		 }
		\label{fig:TiSe2}
	\end{figure}

	This section illustrates the steps of the \xtec\ pipeline benchmarked on a well-known CDW material: TiSe$_2$ \cite{DiSalvo:1976bc, Wilson:1969dv}.  Fig~\ref{fig:TiSe2}   shows the outcome of \xtec d applied to XRD data of bulk 1T-TiSe$_2$, collected at the Cornell High Energy Synchrotron Source (CHESS). As a test case, we specifically explored non-Bragg trajectories with the number of clusters set to $K=2$.

	\xtec\ starts by collecting XRD data on a single crystal encompassing many Brillouin zones in reciprocal space over a range of temperatures $\{T_1, \cdots, T_{d^T}\}$ [see Fig.~\ref{fig:TiSe2}(a)]. The data is then put through a two-stage preprocessing to deal with two key challenges against working with comprehensive data: the volume and the dynamic range of the intensity scale. First, we threshold our data in order to simultaneously reduce its size and isolate its meaningful features. The volume is set by the  $\sim 10^9$ grid points in 3D reciprocal space grid $\{\vec{q}=(q_x,q_y,q_z)\}$ for a single temperature and the 10-30 temperature measurements typically collected.
	However, the relevant peaks are sparse in $\vec{q}$-space for crystalline samples. We thus developed an 
	automated thresholding algorithm (SM section II.B) which removes low intensity noise and reduces the number of $\vec{q}$-space points to be canvassed from the full grid to a selection of points $\{\vec{q}_i\}$, see Fig.~\ref{fig:TiSe2}(b). Removing the intensities associated with Bragg peaks further simplifies the search.

	Second, we rescale the remaining temperature series $\{I(\vec{q}_i, T_j), j=1,\cdots,d^T\}$ still exhibiting a formidable dynamic range [see Fig.~\ref{fig:TiSe2}(d), and Fig.~2(a) of main text] in order to compare trajectories at different scales, focusing on their temperature dependence rather than the absolute scale. Different rescaling schemes can be applied depending on the nature of the data. For TiSe$_2$, since Bragg peak intensities are removed, subtracting each non Bragg intensity trajectory by its temperature mean is sufficient to extract their distinct temperature dependencies. For other data sets that involve Bragg peaks in the analysis such as \csrs, the considerable increase in the range of intensity requires a better rescaling scheme. For this, each intensity trajectory is assigned a z-score (divided by standard deviation after its average value is subtracted). With some datasets such as \srs, we find it useful to employ an alternative rescaling scheme that facilitates further variance-based thresholding as described in the following section (SM. II-B).
	
	[5] We now cluster the resulting collection of preprocessed temperature trajectories,  ${\bf \tilde{I}}({\vec{q}_i})\equiv\{\tilde{I}(\vec{q}_i, T_j); j=1,\cdots,d^T\}$ for each $\vec{q}_i$ to discover qualitatively distinct types of temperature dependence in the data. 
	For this, we adopt a Gaussian mixture model (GMM) \cite{Murphy2013}. Our approach with \xtec d is to ignore correlations between different reciprocal space points ($\vec{q}'s$) and treat each temperature series ${\bf \tilde{I}}({\vec{q}_i})$ as an independent point in the $d^T$ dimensional Euclidean space $\mathbf{R}^{d^T}$.  The GMM assumes that each point in the data set $\{{\bf \tilde I}(\vec{q}_i)\}$ has been independently and identically generated by a weighted sum of $K$ distinct multivariate normal distributions. The number of clusters, $K$, is the only parameter we set manually.  
	
	From Fig.~\ref{fig:TiSe2}(e),  the contrast between the means of the yellow cluster and the teal cluster makes it evident that the yellow cluster represents the order parameter and the temperature at which it crashes down is the critical temperature. 
	The separation between the means exceeding the individual variance affirms the clustering to be a meaningful result. Interpretation of the \xtec\ results is immediate upon locating the two clusters in reciprocal space, as shown in Fig~\ref{fig:TiSe2}(c),  and inspecting the raw data. The locations of the yellow pixels identify the CDW wave vector to be $\vec{Q}_{CDW}=\{(\pi, 0, \pi), (\pi,\pi,\pi))\}$, and equivalent momenta in the hexagonal basis. \xtec\ thus detected the CDW transition with the correct transition temperature $T_c=200$ K and correct ordering wavevector $\vec{Q}_{CDW}$\cite{Waszczak1976} without any prior knowledge. 
	
	\subsection{Preprocessing}\label{preprocessing}
	
	In this section we describe the technical details of \xtec.
	A signature difficulty in the analysis of X-ray diffraction data is the existence of physics at several different intensity scales. This is only further exacerbated when probing low-intensity features where the signal-to-noise ratio can be small.
	If one is to employ thresholding as part of some preprocessing, it is imperative to be careful in order to avoid thresholding-out any important physics. Nevertheless, thresholding is extremely useful for mitigating the influence of noise and for reducing dataset size since most single crystal x-ray diffraction patterns are sparse. Consequently, we propose a new thresholding methodology for isolating the physically relevant regions of k-space.
	
	A naive way to cluster the type of datasets offered by single crystal x-ray diffraction is to apply an i.i.d. assumption and directly try to cluster the associated trajectories, $I(\vec{q}_i,T)$, so that each $q$-point is classified according to its functional temperature dependence. However, such an attempt is immediately thwarted by the existence of a continuum of trajectories spanning over a large intensity range as seen in Fig.~2(a) so that getting any meaningful clustering is difficult. The standard way of dealing with this is to use feature scaling a.k.a. standardization in which one removes the mean for each trajectory and then normalizes it by dividing by its standard deviation.  However, the dominant features of x-ray diffraction data are usually relatively well-localized peaks and most trajectories may be attributed to background fluctuations and thermal diffuse scattering. These trajectories have small, finite means and variances so that conventional standardization amplifies the underlying experimental error and noise, thereby spoiling any immediate attempt at clustering.  On the other hand, failing to standardize makes it difficult to cluster over different energy scales since low-intensity variations can be washed out by larger ones. Thus some cutoff is needed in order to avoid clustering over noise while maintaining the ability to cluster over different energy scales.
	
\begin{figure}[H]
		\centering
		\includegraphics[width=0.7\linewidth]{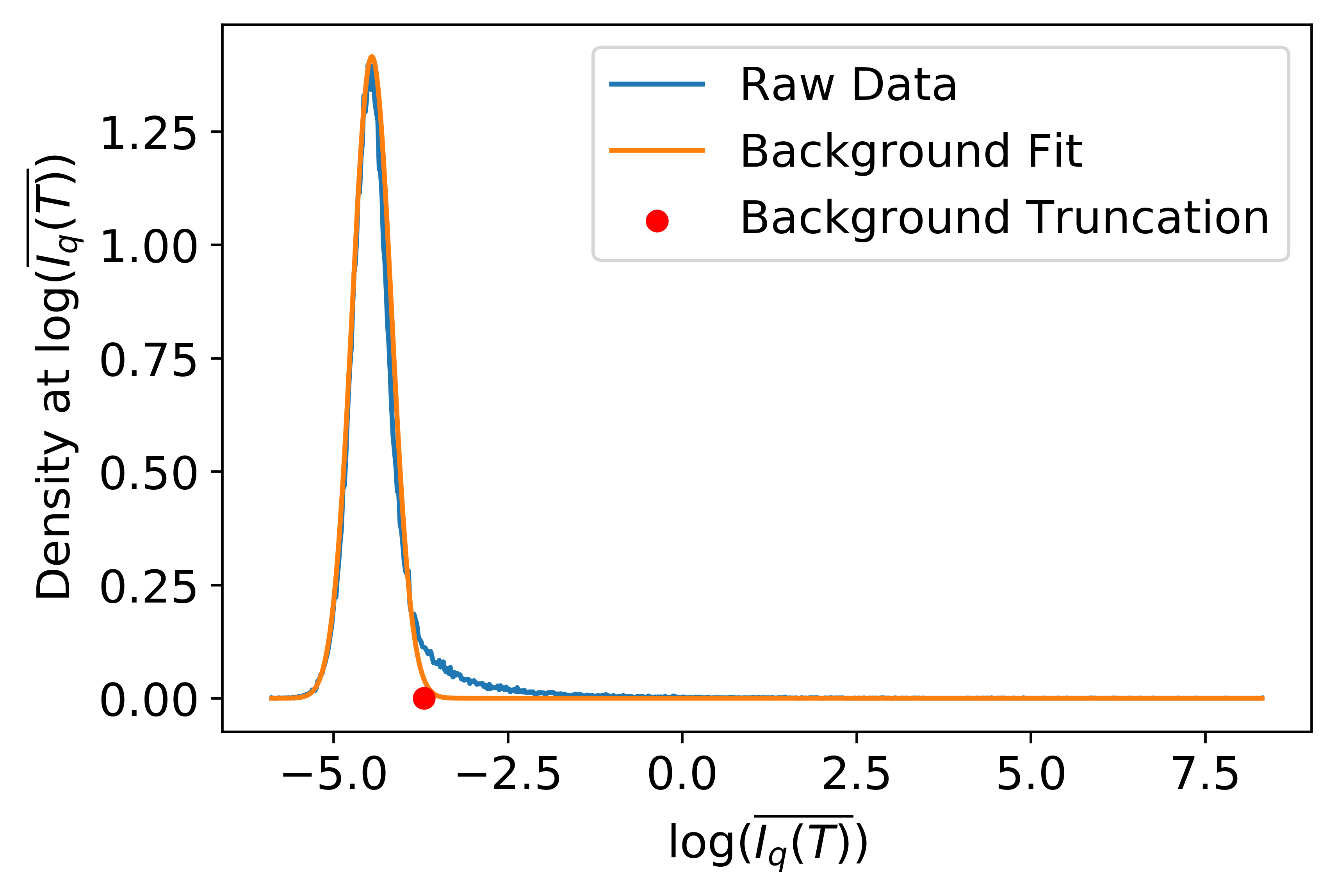}
		\caption{Histogram (blue) of $\log \overline{I(\vec{q}_i,T)}$ for a single unit-cell of TiSe$_2$ with background fit (orange) and truncation point described in the main text.}
		\label{fig:KL-threshold}
	\end{figure}

	In order to properly threshold our data, we exploit the statistical properties of our trajectories'  average intensities, $\log \overline{I(\vec{q}_i,T)}$. Here, the average is performed over temperature so that a single average intensity is obtained for each $\mathbf{q}$.  Several properties of our data make it advantageous to examine the statistics of $\log \overline{I(\vec{q}_i,T)}$ rather than $\overline{I(\vec{q}_i,T)}$, most notably its positive semi-definiteness and large range. Since the dominant features our data are naturally sparse and the background trajectories are characterized by possessing small means and variances, we should expect the distribution of $\overline{I(\vec{q}_i,T)}$ to be sharply peaked near some relatively small background value. Looking at the logarithm,  $\log \overline{I(\vec{q}_i,T)}$, broadens this peak allowing us to resolve the finer structural details of this low-intensity background.
	To first order, we find the distribution of $\log \overline{I(\vec{q}_i,T)}$ to be well-characterized by a bulk background contribution that is approximately normally distributed at low intensities with sparsely distributed high intensity contributions. This can be seen in when looking at the distribution of $\log \overline{I(\vec{q}_i,T)}$ for a single unit-cell of TiSe$_2$ in fig. \ref{fig:KL-threshold}. In order to separate these high intensity features from rest of the data, we take advantage of their sparsity relative to the background. Specifically, we minimize the Kullback-Leibler divergence $D_{KL}$, where for probability
	distributions $p(x),q(x)$: 
	
	\begin{equation}
	D_{KL}( p(x) || q(x) ) = \sum\limits_{x \in X} p(x) \ln \frac{p(x)}{q(x)}
	\end{equation}
	
	\noindent between the distribution of $\log \Big( \overline{I(\vec{q}_i,T)} \Big)$ with a high intensity cutoff and a gaussian. Information theoretically, the Kullback-Leibler divergence quantifies the information loss associated with approximating the distribution $p(x)$ by $q(x)$. In this context, the minimizing $D_{KL}$ optimally chooses a high-intensity cutoff so that the distribution of the remaining $\log \overline{I_{\mathbf{q}} (T)}$ looks closest to a normal distribution. This is illustrated by applying our procedure to a single unit-cell of TiSe$_2$ in Fig. \ref{fig:KL-threshold}. Optimization is performed via gradient descent. Note that optimizing with this sliding cutoff is necessary and a Gaussian cannot be directly fitted because the distribution $\log \overline{I(\vec{q}_i,T)}$ is heavy tailed. Directly fitting with a Gaussian yields a higher cutoff susceptible to missing important low-intensity features.
	
	After thresholding, we find it convenient for the (Ca$_x$Sr$_{1-x}$)$_3$Rh$_4$Sn$_{13}$ data to rescale by dividing by the mean and subtracting one. 
	This has the advantage over z-scoring of allowing us to implement another thresholding step in which we only cluster over high variance trajectories. In particular, it bolsters the model's ability to cluster distinct functional behaviors together because clusters can no longer be smoothly connected to the origin. However, in the case of the TiSe$_2$, we found it sufficient to simply subtract the mean.
	
	\subsection{Label Smoothing}
	{\it X-TEC}'s i.i.d. assumption ignores correlation between nearby momenta and between different Brillouin zones. {\it X-TECs}
	incorporate label smoothing as a first order approach for incorporating these correlations by allowing labels to diffuse between neighboring points and between unit cells. It ultimately results in cleaner, smoother classifications that better align with physicists intuition for order parameters.
	

	Typical label smoothing is a semi-supervised method in which there exists a ground truth for certain points.  These labelings are then ``clamped" and diffused through the rest of the system. Here, we lack a bona fide ground truth and so instead incorporate label smoothing dynamically in between the E and M steps of our EM algorithm. Physically, this adds a diffusive ``force" to our update scheme that encourages a similar labeling of nearby points and points differing by a reciprocal lattice vector. Convergence in this modified EM method occurs when an equilibrium is reached between this diffusion and the GMM clustering.
	
\begin{figure}[H]
		\centering
		\includegraphics[width=0.5\linewidth]{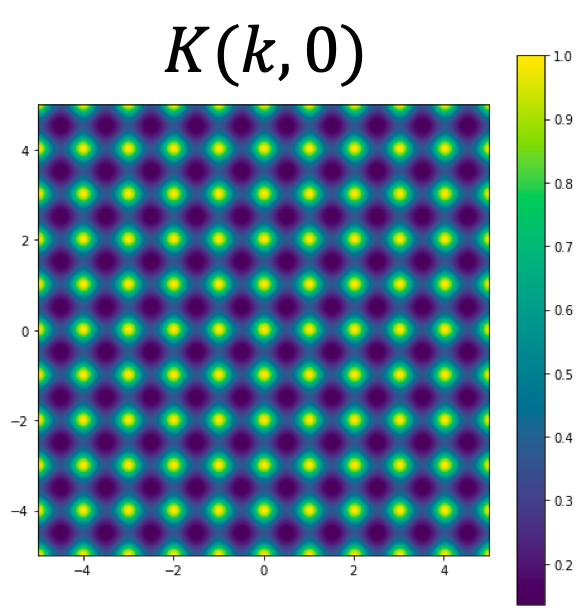}
		\caption{Kernel, $K(k, 0)$, showing the similarity between the origin and momenta in a 2D. }
		\label{fig:kernel}
	\end{figure}

	Our label smoothing requires us to construct a weighted graph connecting similar momenta in order for diffusion to occur. This may be done by computing the following kernel:
	
	\begin{equation}
	K(k, k') = \exp\Bigg[-\sum\limits_{i}\sin^2\Bigg(\frac{Q_i}{2} \cdot (k - k')\Bigg)/\ell^2 \Bigg]
	\label{eqn:labeling_kernel}
	\end{equation}
	
	\noindent where the $Q_i$ are the reciprocal basis vectors and $\ell$ is the relevant length scale for the local correlations. The structure of this kernel is shown in Fig. \ref{fig:kernel} where $K(k, 0)$ is plotted as an intensity for a 2D slice. 
	
	This kernel is really just a weighted adjacency matrix. By incorporating a cutoff in the weights, we may exploit the sparsity of our system for fast matrix-vector multiplication. When handling large datasets, this cutoff is essential since the full kernel is too large to be stored in any reasonable amount of RAM. Define $A$ to be the matrix associated with this kernel after having normalized the rows i.e. it is row stochastic so that $\sum\limits_j A_{i j} = 1$. Now define $P$ to be the matrix consisting of cluster probabilities calculated by the E-step. Specifically, let the first index correspond to the different momenta and the second to the cluster probabilities so that $P$ is also row stochastic. Then the product $A P$ is also row stochastic since $\sum\limits_{j k} A_{i j} P_{j k} = \sum\limits_{j} A_{i j} (1) = 1$. So by multiplying $P$ by A, we generate a new set of diffused cluster probabilities. The strength of this diffusion can be controlled by the number of matrix multiplications. However, note that we cannot simply apply $A$ until $A^n P$ converges, because the largest eigenvector of A is just the constant vector. In practice, we find that even a single application of $A$ between E- and M-steps is sufficient for obtaining smooth labelings. 
	
	\subsection{\xtec d analysis of \srs}
\begin{figure}[H]
		\centering
		\includegraphics[width=0.8\linewidth]{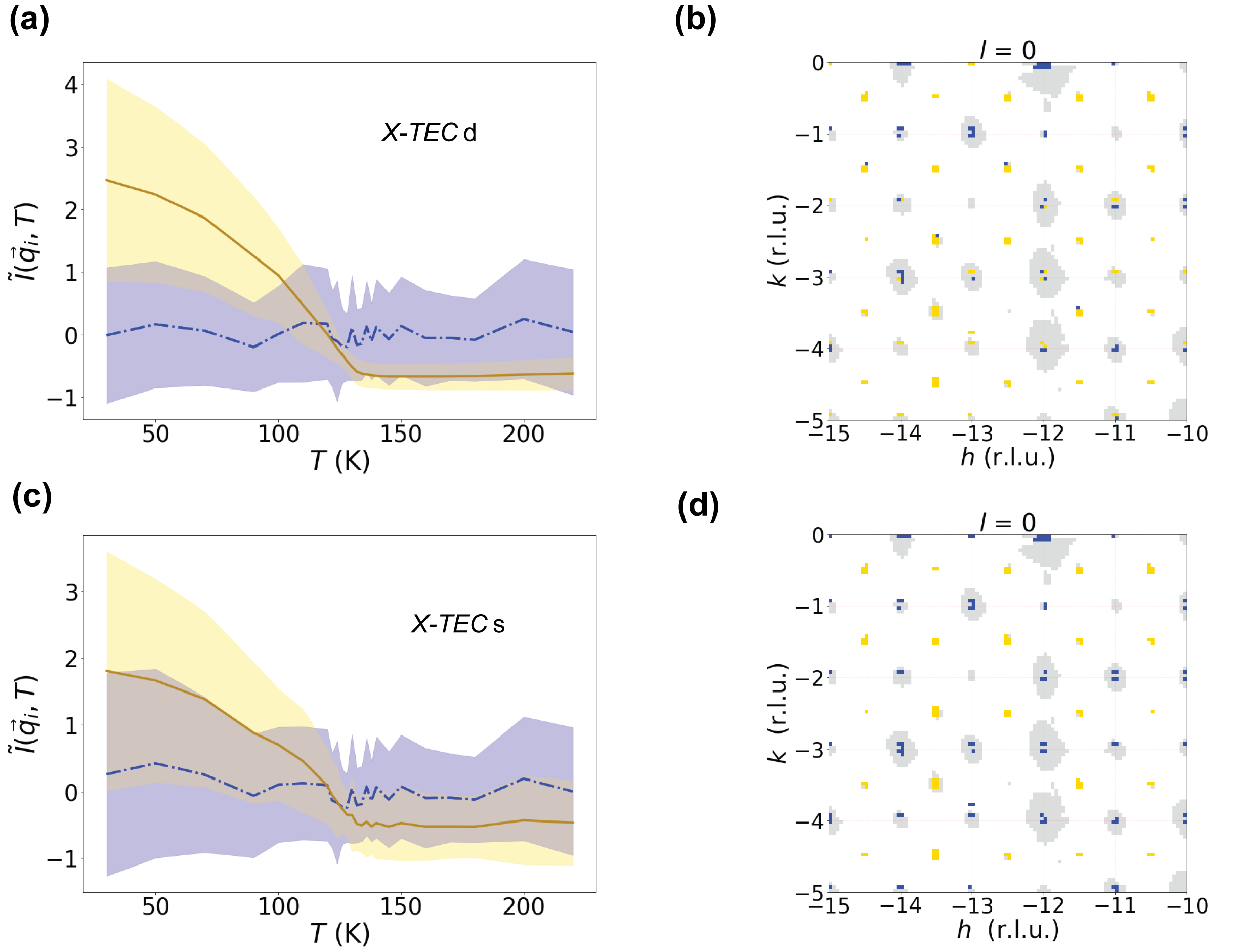}
		\caption{Contrasting the unsmoothed  \xtec d results of upper panels (a-b) with the label smoothed \xtec s results  of lower panel (c-d) for the XRD data from \srs.  {\bf (a,c)}  The cluster trajectories with clustering assignments color-coded as yellow and blue.  $\tilde{I}(\vec{q}_i,T)$ are the rescaled intensities by dividing each intensity trajectory with its mean over temperature  and subtracting one. The lines represent cluster means and the shaded region shows one standard deviation.  {\bf(b,d)} The corresponding yellow/blue cluster assignments in the $(h,k,0)$ plane. The low intensity background (white) and   $\tilde{I}(\vec{q}_i,T)$ with low temperature variance (grey pixels) are excluded from GMM clustering by the preprocessing.}
		\label{fig:smooth_unsmooth}
	\end{figure}

	In Fig.~2 of the main text, we report the \xtec s results of \csrs\ XRD data. Here we contrast the \xtec s results of \srs\ with the unsmoothed \xtec d analysis in Fig.~\ref{fig:smooth_unsmooth}. Without label smoothing,  Fig.~\ref{fig:smooth_unsmooth}(b) shows that nearby $\vec{q}_i$ points are often assigned to different clusters. Label smoothing automatically harmonizes the assignments in the vicinity of each peak at the cost of weakening the cluster separation.
	
	\subsection{Derivation of EM algorithm for GMM and general proof of convergence.}
	\label{label:EM_GMM}
	We follow derivations in Refs. \citenum{Murphy2013} and \citenum{ng_lecturenotes}.
	First recall Jenson's inequality: for convex function f and random variable X, $\mathbb{E}[f(X)] \ge f(\mathbb{E}[X])$ where for strictly convex functions, equality holds iff $X = \mathbb{E}[X]$ almost surely. Let $\ell(\theta)$, denote the model log-likelihood and $X$ be our dataset with $x_i \in X$. Then
	
	\begin{align}
	\begin{split}
	\ell(\theta) = \log p(X ; \theta) = \sum\limits_i \log p(x_i ; \theta) = \sum\limits_i \log \sum\limits_{z_i} p(x_i, z_i ; \theta) \\
	= \sum\limits_i \log \sum\limits_{z_i} q_i(z_i) \frac{p(x_i, z_i ; \theta)}{q_i(z_i)} \ge \sum\limits_{i, z_i} q_i(z_i) \log \frac{p(x_i, z_i ; \theta)}{q_i(z_i)} \equiv \tilde{\ell}(q,\theta)
	\end{split}
	\end{align}
	
	\noindent where $q_i(z_i)$ is some distribution over a random variable $z_i$ (in our case this will be the cluster assignment) s.t. $\sum\limits_{z_i} q_i(z_i) = 1$ and we have used Jenson's inequality. In order for this bound to be tight, $X = \mathbb{E}[X] \implies q_i(z_i) = p(z_i | x_i ; \theta)$. Tightness of this bound implies that improving $\tilde{\ell}(q,\theta)$ necessarily improves $\ell(\theta)$ but since theta is unknown, we will have to make a guess, $\theta_t$, and improve it iteratively.  This iterative prescription is known as expectation maximization (EM). It consists of an E-step, where $q_i^t \leftarrow p(z_i | x_i ; \theta_t)$ and an M-step $\theta^{t+1} \leftarrow \argmax\limits_{\theta} \tilde{\ell}(q^t,\theta)$.
	
	We now derive the EM algorithm for the GMM. The E-step follows directly from the model likelihood and Bayes' theorem:
	\begin{align}
	\begin{split}
	w_i^k &\equiv p(z_i = k | x_i; \pi_k, \mu_k, \Sigma_k) = \frac{\pi_k \mathcal{N}(x_i | \mu_k, \Sigma_k)}{\sum\limits_k \pi_k \mathcal{N}(x_i | \mu_k, \Sigma_k)} \\
	\mathcal{N}(x_i | \mu_k, \Sigma_k) &\equiv \frac{1}{(2 \pi)^{n/2}} \frac{1}{\sqrt{\det \Sigma_k}} e^{-\frac{1}{2}(x_i - \mu_k)^{\dagger} \Sigma_k^{-1} (x_i - \mu_k)}
	\end{split}
	\end{align}
	
	For the M-step, we must find $\{ \pi, \mu,\Sigma \}$ that optimizes our lower log-likelihood bound:
	
	\begin{equation}
	\tilde{\ell}(\{w^k_i, \pi_k, \mu_k, \Sigma_k\}) = \sum\limits_{i, k} w_i^k \log\Big[ \frac{\pi_k \mathcal{N}(x_i | \mu_k, \Sigma_k)}{w_i^k} \Big] + \lambda (1 - \sum\limits_k \pi_k)
	\end{equation}
	
	\noindent where $\lambda$ is a Lagrange multiplier constraining the mixing weights to sum to unity.
	
	Solving for the mixing weights:
	
	\begin{align}
	\begin{split}
	0 &=  \partial_{\pi_j} \tilde{\ell} = \sum\limits_{i, k} w_i^k \frac{1}{\pi_k} \delta_{j k} - \lambda \sum\limits_k \delta_{j k} \implies \lambda = \frac{1}{\pi_j} \sum\limits_i w_i^j \\
	\lambda &= \lambda \sum\limits_k \pi_k = \sum\limits_{i,k} w_i^k = \sum\limits_i 1 \equiv m \\
	&\implies \pi_j = \frac{1}{m} \sum\limits_i w_i^j
	\end{split}
	\end{align}
	
	Solving for the mean:
	
	\begin{align}
	\begin{split}
	0 &= \partial_{\mu_l} \tilde{\ell} = 2 \sum\limits_i w_i^l \Sigma_l^{-1} (x_i - \mu_l) \\
	&\implies \mu_l = \frac{1}{\sum\limits_i w_i^l} \sum\limits_i w_i^l x_i
	\end{split}
	\end{align}
	
	Solving for the covariance is a little trickier. First note the following matrix identities for symmetric invertible matrix A:
	
	\begin{align}
	\begin{split}
	&\partial \big(\log (\det A) \big) = \Tr (A^{-1} \partial A) \\
	&\partial A^{-1} = - A^{-1} (\partial A) A^{-1}
	\end{split}
	\end{align}
	
	Now, when solving for the covariance we promote the covariance cluster index to an upper index so that the lower indices refer to the matrix elements:
	
	\begin{align}
	\begin{split}
	0 &= \partial_{\Sigma_{m n}^l} \tilde{\ell} = \sum\limits_{i, k} w_i^l \partial_{\Sigma_{m n}^l} \Big[ \log \det \Sigma^k + (x_i - \mu_k)^{\dagger} (\Sigma^{k})^{-1} (x_i - \mu_k)\Big] \\
	&= \sum\limits_{i, k} w_i^l \Big[ \delta_{l k} \Tr \big\{ (\Sigma^{k -1})_{r s} \delta_{s m} \delta_{t n} \big\} - \delta_{l k}  \sum\limits_{ps} (x_i - \mu_k)^{\dagger}_p \Big\{\sum\limits_{q r} (\Sigma^{k -1})_{p q} \delta_{m q} \delta_{n r} (\Sigma^{k -1})_{r s} \Big\} (x_i - \mu_k)_s \Big] \\
	&= \sum\limits_i w_i^l \Big[ \Sigma^{l - 1}_{n m} - \sum\limits_{p, s} (x - \mu_l)^{\dagger}_p \Sigma^{l -1}_{pm} \Sigma^{l - 1}_{n s} (x - \mu_l)_s \Big] \\
	&= \sum\limits_i w_i^l \Big[ \Sigma^{l - 1} - \Sigma^{l - 1} (x_i - \mu_l) (x_i - \mu_l)^{\dagger} \Sigma^{l - 1} \Big] \\
	0 &= \sum\limits_i w_i^l \Big[ \Sigma^l - (x_i - \mu_l) (x_i - \mu_l)^{\dagger} \Big] \\
	&\implies \Sigma_l = \frac{1}{\sum\limits_i w_i^l} \sum\limits_i w_i^l (x_i - \mu_l)(x_i - \mu_l)^{\dagger}
	\end{split}
	\end{align}
	
	Note that all quantities derived about have the same form as one would expect from standard regression but with each data point $x_i$ having a cluster weight $w_i^k$.
	
	\section{$\rm{Cd}_2\rm{Re}_2\rm{O}_7$ Analysis}
	\subsection{Specific Heat Measurements}
\begin{figure}[H]
		\centering
		\includegraphics[width=0.7\linewidth]{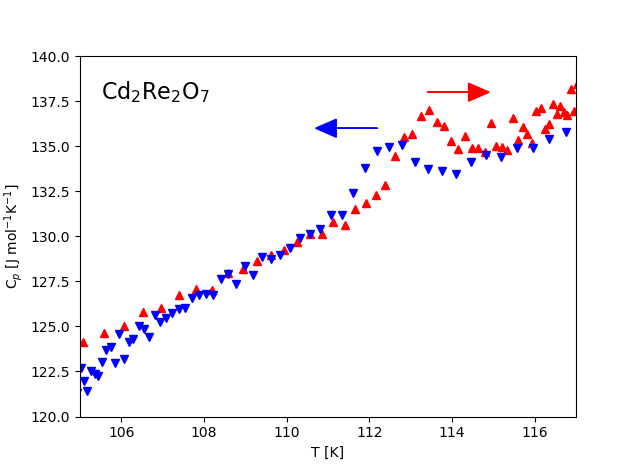}
		\caption{Specific heat of \cro, measured using the scanning method described in the text.}
		\label{fig:Cp_SI}
	\end{figure}

	In the main text, the heat capacity ($C_p$) of \cro\ was displayed in Fig.~3(b). The data shown in that figure was processed by a standard method in relaxation calorimetry (“pseudostatic method”) in which the heat capacity is assumed to be constant throughout the heating and cooling segments of an applied heat pulse during which $\Delta T\lll$T. However, in the presence of a 1st order transition, the shape and magnitude of a peak in $C_p$ at the phase transition temperature can be modified, while the hysteresis can be lost, when using the pseudostatic method.  For this reason, we have also used the “scanning method” for which $C_p$ is numerically determined at every point in the warming and cooling segments, which yields a more accurate peakshape and hysteresis for a 1st order transition at the cost of noise and absolute accuracy. A more detailed description of pseudostatic and scanning analysis can be found in Ref. \citenum{Gillard.2015}. Fig. \ref{fig:Cp_SI} shows the temperature dependence of $C_p$ in the vicinity of the $\sim113$~K phase transition when analyzed using the scanning method. A small but resolvable thermal hysteresis was observed between the peaks in $C_p$ from the heating and cooling curves, which is suggestive of a latent heat and hence a first-order character. We do note, however, that the peak height and width of the peak in $C_p$ did not differ substantially between these two methods, as would also be anticipated for a first-order transition, and for this reason the analysis of $C_p$ alone is not definitive in identifying the order of the transition.
	
		\subsection{Low resolution \cro~ XRD data}
\begin{figure}[H]
		\centering
		\includegraphics[width=0.9\linewidth]{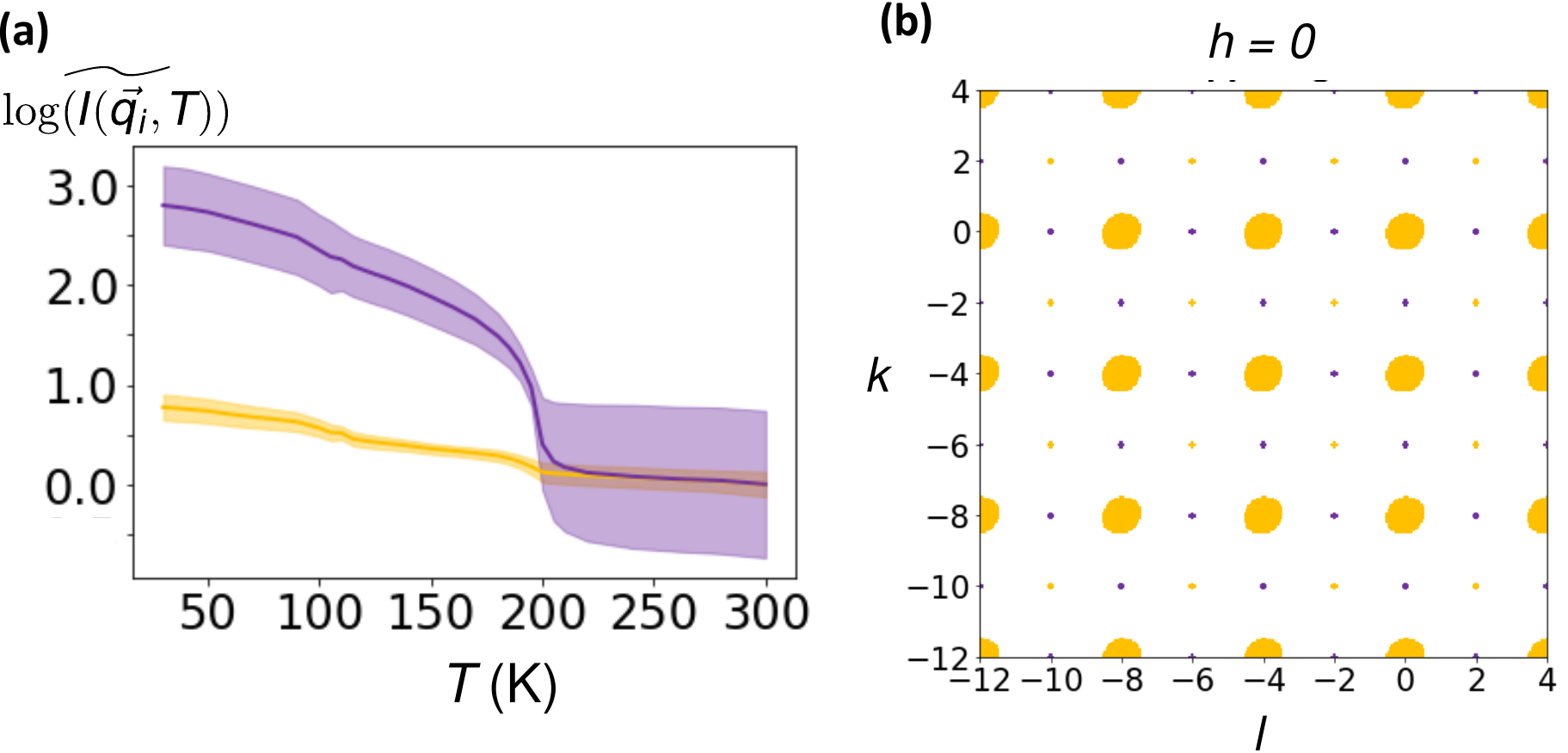}
		\caption{{Two-cluster \xtec s results on the lower resolution data 
				spanning 15,000 BZ's of \cro. \bf (a)} Cluster means (solid lines) and standard deviations (shaded areas) for the two clusters are shown in purple and yellow, interpolated between $d^T=30$ temperature points of measurement. The data is peak averaged prior to the \xtec\   preprocessing to suppress fluctuation signal and isolate the transition at $T_{s2}$. $\widetilde{\log(I(\vec{q}_i,T))}$ 
			denotes rescaled data by logging the peak averaged intensity, and subsequently subtracting its mean over temperature.
			{\bf (b)} The cluster assignements of thresholded $\vec{q}_i$ points that belong to the two clusters in (a) in a portion of the $h = 0$ plane. 
		}
		\label{fig:CRO_low_res}
	\end{figure}

	This section discuss the \xtec\ analysis with a lower resolution XRD data of \cro. We first performed scans using an x-ray energy of 87 keV, which contained scattering spanning nearly 15,000 Brillouin zones, 
	A first pass of \xtec s\footnote{here we simply averaged peaks due to the volume of the data.} for two clusters ($K=2$) readily finds a cluster whose intensity rises sharply below $T_{s1}=200$ K [the purple cluster in Fig.~\ref{fig:CRO_low_res}(a)]. 
	The crisp clustering results with tight variance around the means reflect the amplification of the meaningful trend upon using data from a large number of BZ's. By examining the \xtec\ cluster assignments, we find the purple cluster to exclusively consist of peaks with $\vec{Q}=(H,K,L)$, with all indices even, exactly one of which is not divisible by four, using the cubic indices of Phase I [see Fig.~\ref{fig:CRO_low_res}(b)]. Peaks that are equivalent in the cubic phase have different temperature dependence in Phase II, implying that the sample is untwinned, something that is confirmed by our high-resolution data. This means that the presence of $(00L)$ peaks with $L=4n+2$ below $T_{s1}$ in phase II unambiguously rules out all the tetragonal space groups compatible with the pyrochlore structure, apart from $I\bar{4}m2$ and $I\bar{4}$. According to an earlier group theoretical analysis\cite{Ivan:2003dl}, of these two, only the former is compatible with a single second-order phase transition, so our data is strong confirmation of previous conclusions that, at $T_{s1}$, $I\bar{4}m2$ phase is selected out of two-dimensional $E_u$ representation\cite{petersen:np2006a,yamaura:pr2017a}. 
	
	\subsection{Preprocessing and clustering setup details}
	Here we specify different preprocessing steps and clustering choices for the analysis of $\rm{Cd}_2\rm{Re}_2\rm{O}_7$ high resolution data presented in Fig.~3 and 4 of main text, as well as the lower resolution data in Fig~\ref{fig:CRO_low_res}.
	\begin{itemize}
		
		\item{\bf \xtec s (peak averaged) on cubic forbidden peaks of high resolution data: Fig 3(c), Fig 4(a)}
		
		\begin{enumerate}
			\item We begin by selecting a $50 \times 50 \times 50$ region around each known peak center
			and thresholding as described in SM~\ref{preprocessing}.
			\item We then floodfill from the peak centers and average all resulting trajectories
			to form a single, averaged trajectory per peak. 
			\item We rescale the data by z-scoring it.
			\item We exclude all the cubic allowed Bragg peaks, and restrict the temperature range to [30 K, 150 K] so that \xtec\ can focus on better resolving the distinct cluster trajectories across $T_{s2}$. See Fig~\ref{fig:CRO_peak_averages} for the same analysis, but including all Bragg peaks and over the full temperature range.
			\item We cluster the peak-averaged trajectories using K = 2 clusters.  We found two clusters to be the minimum number necessary to separate all 
			distinct behaviors and that there was no advantage to using more than two.  
			\item The dashed lines in Fig.~3(c) and symbols in Fig 4(a) show the cluster averaged  intensity trajectory of the two clusters. The cluster averaged trajectory is shown for the full temperature range: [30 K, 300 K], although the clustering assignments were obtained from trajectories $\le150$ K.
		\end{enumerate}
		
		\item{\bf \xtec d (peaks opened) on cubic forbidden peaks of high resolution data: Fig 3(c-d), Fig: 4(c)}
		
		\begin{enumerate}
			\item We select a $50 \times 50 \times 50$ window around each known
			peak center and threshold as described in SM II(a).
			\item We only include peaks that are forbidden in the cubic phase. The temperature range is restricted to [30 K, 150 K] like in the peak averaged \xtec s analysis.
			\item We rescale the data by z-scoring it.
			\item We cluster the data using $K = 3$ clusters. \item The resulting cluster averaged intensity trajectory for the full temperature range: [30 K, 300 K] is shown as solid lines in Fig.~3(c).
			\item Cluster averaging the absolute intensity trajectories washes out the characteristic temperature dependence of the diffuse halos. This can be remedied by cluster averaging over z-scored intensities. This is reported in Fig. 4(c). To confirm that z-scored intensity indeed represents the behavior of diffuse halos, see SM Fig.~\ref{fig:CRO_diffuse_scattering } for the absolute intensity trajectories of the diffuse halos in two manually selected peaks.         
		\end{enumerate}
		\item{\bf \xtec s (peak averaged) on low resolution data: \bf Fig~\ref{fig:CRO_low_res}}
		
		\begin{enumerate}
			\item In order to reduce noise, we first construct an average BZ mask by thresholding every BZ as described in SM~\ref{preprocessing} and then averaging the thresholded BZs together. 
			\item We then manually select a cutoff value for the averaged BZ that maintains all 
			the peaks while removing as much background as possible, and set each $\vec{q}$-point 
			in the average BZ with value greater than the cutoff to 1, and the rest to 0 to form the
			mask. 
			\item We multiply each BZ by the average BZ mask to remove noise and emphasize the peaks.
			\item Beginning from the known peak centers, we floodfill to pick out all $\vec{q}$-points
			belonging to each peak.
			\item We perform peak averaging by averaging the trajectories of all $\vec{q}$-points
			belonging to each peak and replacing them with the single, averaged trajectory.
			\item We rescale the data by taking the log of one plus each peak-averaged trajectory,
			and subsequently subtracting the mean.
			\item Finally we cluster using K = 2 clusters. We subtract the minimum value
			of the cluster means when plotting to emphasize the order-parameter like behavior
			of the purple cluster in Fig.~\ref{fig:CRO_low_res}(a).  Here \xtec\ analyses the data for the full temperature range [30 K, 300 K]. 
		\end{enumerate}
		
	\end{itemize}
	
	\subsection{Structure Factor Analysis}

\begin{figure}[H]
		\centering
		\includegraphics[width=0.7\linewidth]{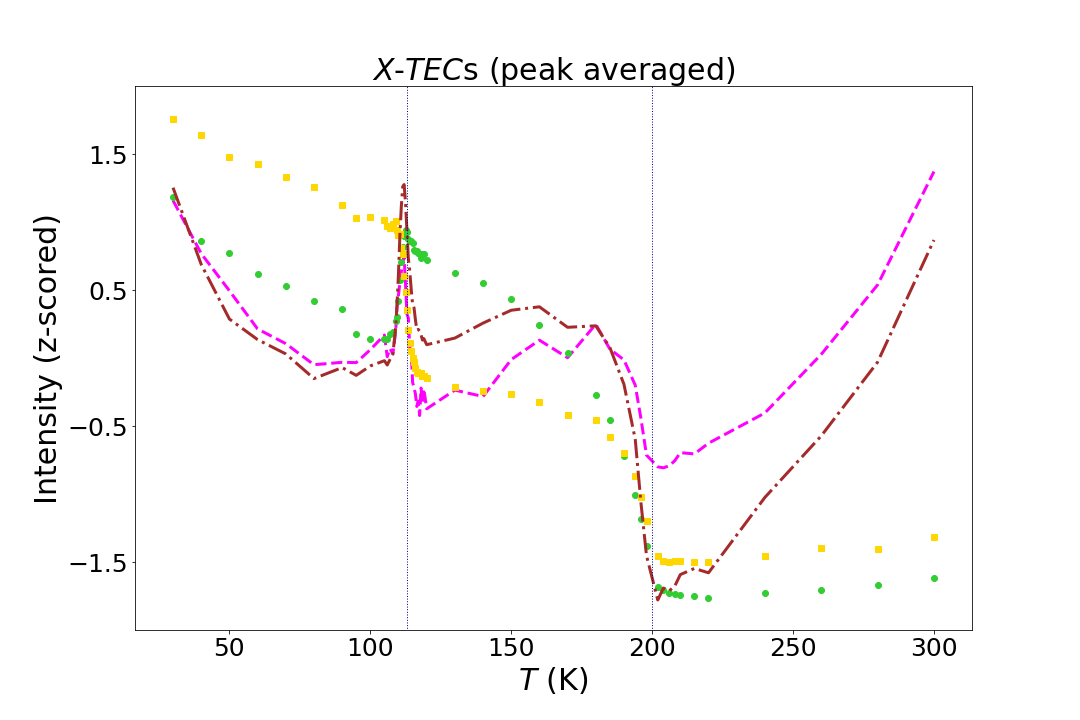}
		\caption{Four-cluster \xtec s (peak averaged) results on the high resolution measurements of \cro\ retaining all Bragg peaks. Two of these sub-cluster trajectories (yellow and green symbols) identify with the cubic forbidden trajectories shown in Fig.~3(c) and Fig.~4(a) of main text. The other two sub-cluster trajectories (magenta and brown lines) arise from peaks that are not forbidden in the high-temperature cubic phase. The temperatures of the two structural phase transitions are shown as dotted lines.}
		\label{fig:CRO_peak_averages}
	\end{figure}

    Fig~3 and 4 of main text discuss the two-clustering \xtec s  analysis after excluding the cubic allowed peaks. By including all Bragg peaks, four clusters are sufficient for \xtec s to identify all the distinct trajectories. Fig. \ref{fig:CRO_peak_averages} shows the cluster means (z-scored intensity trajectories) for all four clusters identified by \xtec s (peak averaged) analysis. Two of these sub-clusters (yellow and green symbols) can be identified with the behavior of the cubic excluded peaks. It should be noted that these clusters represent the average temperature dependence of all the peaks assigned to their respective clusters, so there can be large variations within each cluster. However, the ML analysis has identified distinctive behavior in each cluster that we have verified by manual inspection of a number of peaks. All four clusters show similar temperature dependence close to the transition at $T_{s1}=200$~K, but strikingly different behavior at the lower transition at $T_{s2}=113$~K. The yellow cluster trajectory show a sudden drop while the green cluster peaks show a sudden increase in intensity across $T_{s2}$. The magenta and brown lines show a sharp spike in intensity at $T_{s2}$, before falling back to their values just above the transition. We do not currently have an explanation for this remarkable behavior. 
	
	The structural phase transition at $T_{s1}$ is from the cubic pyrochlore structure, with space group $Fd\bar{3}m$, to a distorted tetragonal structure, with space group $I\bar{4}m2$. This space group allows distortions of the cadmium and rhenium cations along the $z$ direction and either the $x$ or $y$ direction depending on the Wyckoff positions, using the $I\bar{4}m2$ unit cell, which is rotated by 45\degree\ from the cubic unit cell, \textit{i.e.}, $x$ is parallel to the (110) direction of the high-temperature cubic structure. There are associated displacements of the oxygen ions, but the x-ray measurements are not sensitive to them.
	
	Analytic calculations of the structure factors for the Bragg peaks in terms of the allowed $x$ and $z$ distortions fall into four groups that correspond well to the four ML clusters. For example, the two groups whose intensities are forbidden in the high-temperature cubic phase (yellow and green) have the following form (H,K,L in following equations are in tetragonal indices):
	\begin{equation}
	F_1(H,K,L) \propto (-1)^{n_3} \sum_{M=\mathrm{Cd,Re}}\left\{f_{M}\left[(-1)^{n_1}cos(2\pi H\delta x_{M}) e^{-2\pi i L\delta z_{M}} - (-1)^{n_2}cos(2\pi K\delta x_{M}) e^{2\pi i L\delta z_{M}}\right] \right\}
	\end{equation}
	where $n_1=\frac{1}{2}H$, $n_2=\frac{1}{2}K$, and $n_3=\frac{1}{4}(L-2)$.
	\begin{equation}
	F_2(H,K,L) \propto (-1)^{n_3} \sum_{M=\mathrm{Cd,Re}}\left\{f_{M}\left[(-1)^{n_1}sin(2\pi H\delta x_{M}) e^{-2\pi i L\delta z_{M}} + (-1)^{n_2}sin(2\pi K\delta x_{M}) e^{2\pi i L\delta z_{M}}\right] \right\}
	\end{equation}
	where $n_1=\frac{1}{2}(H-1)$, $n_2=\frac{1}{2}(K-1)$, and $n_3=\frac{1}{4}L$.
	
	It can be seen that, for small values of $H$ and $K$, $F_1(H,K,L)$ are mostly sensitive to distortions along the $z$-axis, whereas for small values of $L$, $F_2(H,K,L)$ is mostly sensitive to in-plane distortions along $x$ or $y$ (where $\delta x=\delta y$). The assignments of individual peaks in the \xtec\ analysis show that the (H,K,L) values of the green cluster are indeed dominated by in-plane distortions whereas the yellow cluster peaks are dominated by $z$-axis distortions. This suggests that the distinctive temperature dependences of peaks in the green and yellow clusters can be used to derive information about the relative distortions along $x$ and $z$. If we assume that the temperature dependence of $\delta x$ and $\delta z$ follows that expected for an order parameter with a common critical exponent, $\beta$, from 200K down to 120K, the peak intensities would vary as $(T-T_c)^{2\beta}$. 
	
	\begin{figure}[H]
		\centering
		\includegraphics[width=0.7\linewidth]{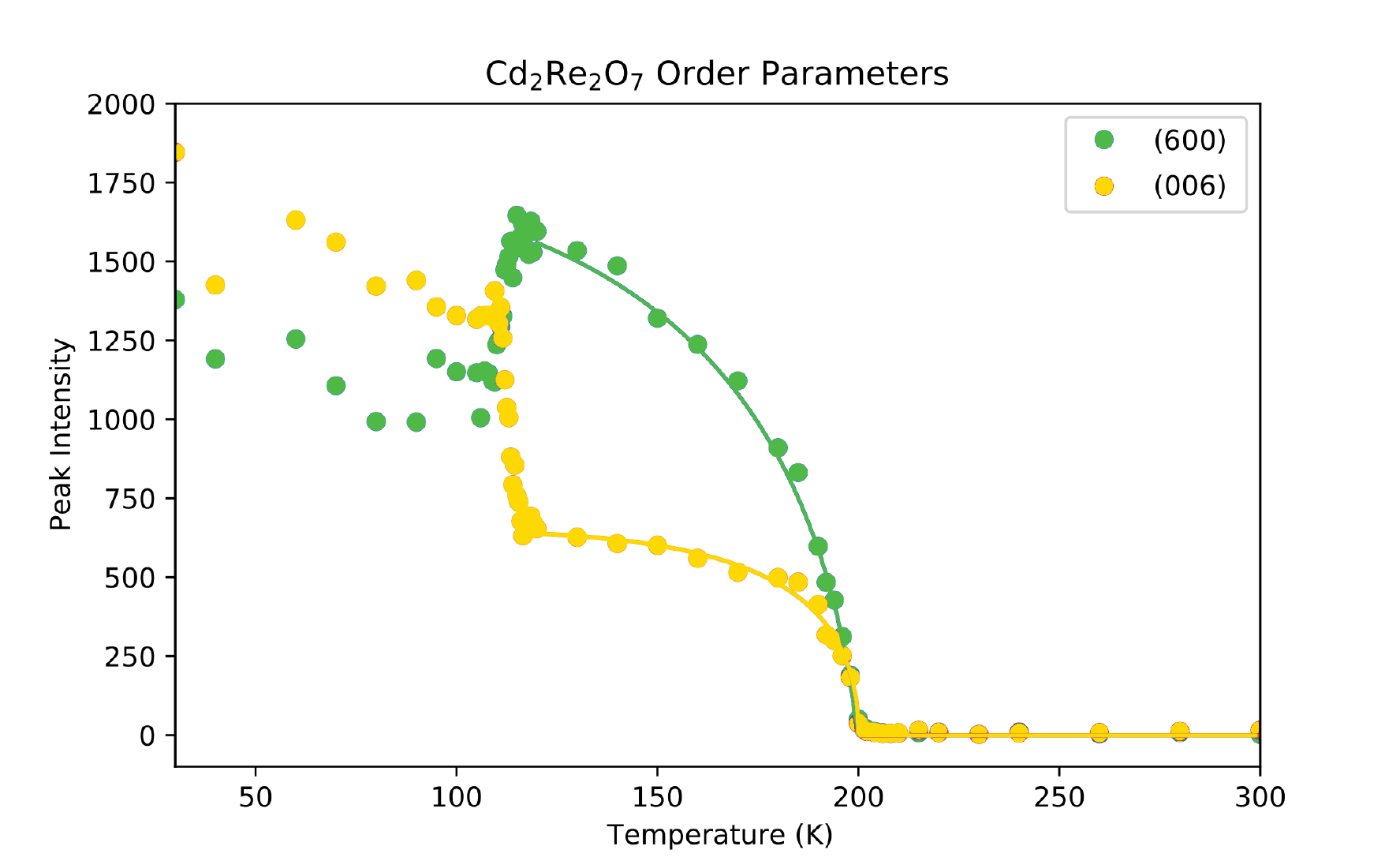}
		\caption{Temperature dependence of the 006 (yellow) and 600 (3$\overline{3}$0) in tetragonal indices) (green) Bragg peaks using the indices of the high-temperature cubic phase. The green and yellow solid lines are fits between 120~K and 300~K to the structure factors in equations 10 and 11, respectively, assuming that the distortions, $\delta x$ and $\delta z$ for the Cd and Re ions, vary as $(T-T_c)^{\beta}$, with $\beta=0.25$ and $T_c=200$~K.}
		\label{fig:CRO_order_parameters }
	\end{figure}
	
	As an example, Figure \ref{fig:CRO_order_parameters } compares the 006 and 060 Bragg peaks using the indices of the cubic phase. These are the peaks that have been assigned to the yellow and green clusters of \xtec s shown in Fig. 4(a). Equations 10 and 11 show that the 006 (yellow) peak is only sensitive to $\delta z_\mathrm{Cd}$ and $\delta z_\mathrm{Re}$, whereas the 060 (green) peak is only sensitive to in-plane distortions. The fit to the 006 peak yields relative $z$-axis distortions that are equal and opposite, \textit{i.e}, $\delta z_\mathrm{Re}=-\delta z_\mathrm{Cd}$, illustrated in Fig. 4(b). The out-of-phase distortions are the reason for the flattening of the peak intensity of the 006 peak between 180~K and 120~K, confirming the conclusions based on the fits to the cluster means in Fig. 4(a). On the other hand, the 060 peak follows the scaling law from 200~K to 120~K, showing either that $\delta x_\mathrm{Re}$ has the same sign as $\delta x_\mathrm{Cd}$ or that one of the distortions is much larger than the other. This is an example where the temperature dependence of the peak intensities below a structural phase transition yields information on the relative internal distortions, which have proved to be too subtle for conventional crystallographic refinement until now.
	
	\subsection{Temperature Dependence of Diffuse Scattering}
\begin{figure}[H]
		\centering
		\includegraphics[width=0.7\linewidth]{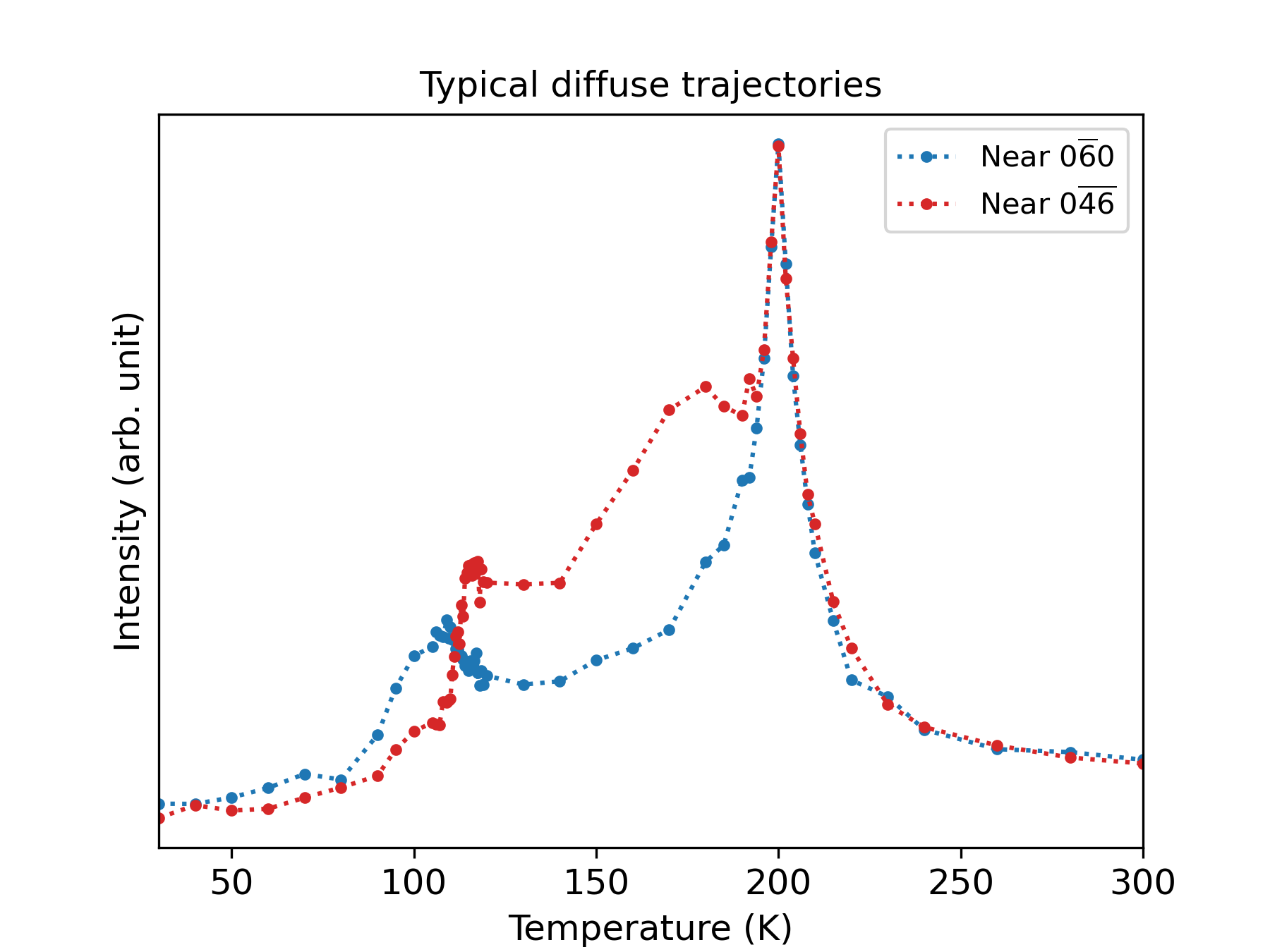}
		\caption{Temperature dependence of the diffuse scattering near the $0\overline{6}0$ and $0\overline{46}$ Bragg peaks shown in Fig. 3 (e-f) (blue and red circles, respectively). The temperature dependences are taken from 0.02x0.02x0.02 rlu bins centered at $hkl = (-0.05, -6.0, 0.05)$ and $hkl = (-6.0, -3.95, -0.05)$.}
		\label{fig:CRO_diffuse_scattering }
	\end{figure}

	Fig. 3 in the main article showed two-cluster $(K=2)$ assignments from \xtec d analysis of the high-resolution data with cubic excluded peaks. This reveals differences between clusters in the diffuse scattering halo around the Bragg peaks, which represent fluctuations in the order parameter at the $\Gamma$ point. Peaks in the blue cluster, which are also assigned to the blue cluster in Fig. \ref{fig:CRO_order_parameters }, display weak diffuse scattering halos in the range $T_{s2}<T<T_{s1}$  while peaks in the red cluster displayed much stronger diffuse scattering halos in this region, as seen in Fig. 3(c). This is clearly illustrated in Fig. \ref{fig:CRO_diffuse_scattering }, where the temperature dependence of the diffuse scattering near the $0\overline{6}0$ and $0\overline{46}$ Bragg peaks are compared. Both show strong critical scattering at $T_{s1}$, but the diffuse contribution is much stronger in peaks assigned to the red cluster, which are most sensitive to the out-of-phase $z$-axis fluctuations of the Re and Cd sublattices. We attribute these strong $z$-axis fluctuations to the Goldstone modes seen in Raman scattering \cite{kendziora:prl2005a}, which are fluctuations between the two nearly degenerate $E_u$ modes. This interpretation is justified by the fact that the Goldstone modes are fluctuations from the $I\bar{4}m2$ ground state to $I4_122$ symmetry, in which $\delta z$ is constrained to be 0. It is therefore not surprising that $z$-axis fluctuations are dominant. 
	
	The \xtec\ trajectories shown in Fig. 4(c) are the cluster average from z-scored intensities. The red cluster shows a substantial increase in diffuse scattering just above $T_{s2}$, whereas the blue cluster shows a weaker peak just below $T_{s2}$. Although 4(c) shows the z-scored intensities and not their absolute values, note its similarity with Fig. \ref{fig:CRO_diffuse_scattering }. This indicates that the \xtec d cluster averages of z-scored intensities give the typical diffuse  trajectories.  
	
	\subsection{Mode energies and intensities from Landau theory}
	
	The Landau free energy in an $E_u$ model for Cd$_2$Re$_2$O$_7$ is \cite{Ivan:2003dl}
	\begin{equation}
	F = a_1Q^2 +a_2Q^4 +a_3Q^6 +a_4Q^8 +Q^6[b_1+cQ^2]\frac{1}{2}[1+\cos(6\phi)]
	\end{equation}
	with $Q$ the order parameter amplitude and $\phi$ the phase angle.  For $I\bar{4}m2$, $\phi$ = 30(2n+1) and for $I4_122$, $\phi$ = 60n (the different angles represent different domains).  $F$ vanishes at $T_{s1}$ and the anisotropy (last term) would vanish at $T_{s2}$ if it were not for the first-order jump in $Q$.  The Goldstone (phase) mode energy is given by \cite{meier:pr2020a}
	\begin{equation}
	\omega_G^2 = \chi_G^{-1}/M
	\end{equation}
	where the inverse Goldstone susceptibility is
	\begin{equation}
	\chi_G^{-1} = \frac{1}{Q^2} \frac{\partial^2F}{\partial\phi^2} = 18Q^4|b_1+cQ^2|
	\end{equation}
	with $M$ some ion effective mass, where in the second expression we have taken into account the value of $\phi$ in the two phases (which leads to the modulus).  The Higgs mode energy is given by
	\begin{equation}
	\omega_H^2 = \chi_H^{-1}/M
	\end{equation}
	where
	\begin{equation}
	\chi_H^{-1} = \frac{\partial^2F}{\partial Q^2} = 2a_1 +12a_2Q^2 +30a_3Q^4 +56a_4Q^6 +30Q^4[b_1+\frac{28}{15}cQ^2]\frac{1}{2}[1+\cos(6\phi)]
	\end{equation}
	The first-order transition from $I\bar{4}m2$ (phase II) to $I4_122$ (phase III) is given by the condition \cite{Ivan:2003dl}
	\begin{equation}
	a_1 = 2a_2(b_1/c) -3a_3(b_1/c)^2 +(4a_4+c/2)(b_1/c)^3
	\end{equation}
	$Q^2$ is given by the cubic equation ($\frac{\partial F}{\partial Q}=0$)
	\begin{equation}
	-2a_1 = 4a_2Q^2 +6a_3Q^4 +8a_4Q^6 + 6Q^4[b_1+\frac{4}{3}cQ^2]\frac{1}{2}[1+\cos(6\phi)]
	\end{equation}
	Finally, the soft mode energy above $T_{s1}$ ($\omega_s$) is gotten by setting $Q$=0 in $\chi_H^{-1}$.  In practice, the effective $M$ is unknown (involving Cd, Nb and O masses), so all mode energies will be multiplied by the same constant in order to agree with Raman data \cite{kendziora:prl2005a} for the Higgs energy at $T$=0 (85 cm$^{-1}$).
	
	We now have all we need to calculate the order parameter, the phase boundary, and mode energies.  What about the mode intensities?  The basic idea can be seen from the work of Fleury \cite{Fleury} and Shapiro \cite{Shapiro}.  The energy integrated intensity (appropriate for the diffuse scattering collected from high energy x-rays) is given by \cite{Shapiro}
	\begin{equation}
	I_q = \frac{1}{\pi} \int [n(\omega)+1] \Im[\omega_q^2 - \omega^2 - i\omega\Gamma_q]^{-1} d\omega
	\end{equation}
	where $n(\omega)$ is the Bose factor, $\omega_q$ is the mode energy for a given $q$, and $\Gamma_q$ is the lifetime broadening.  Assuming we can replace $n(\omega)$ by $T/\omega$, this integral reduces to
	\begin{equation}
	I_q = T/\omega_q^2
	\end{equation}
	This expression is obviously divergent for $q$=0 at $T_{s1}$. To correct for this, we recognize that the data are collected over a small range in $q$.  We assume the $q$ dependence of the mode energy goes like
	\begin{equation}
	\omega_q^2 = \omega_0^2+\alpha^2 q^2
	\end{equation}
	where $\alpha$ results from the gradient terms in the Landau energy.  Integrating over $q$, we obtain
	\begin{equation}
	T \int \frac{q^2dq}{\omega_q^2} \propto T[1-\tilde{\omega}_0\tan^{-1}{\frac{1}{\tilde{\omega}_0}}]
	\end{equation}
	where $\tilde{\omega}_0=\frac{\omega_0}{\alpha q_c}$ with $q_c$ the momentum cut-off.  Since $\alpha$ is unknown (no mode dispersions have been measured for this material), we set $\alpha q_c$ to the lower bound of the Raman data (6 cm$^{-1}$ \cite{kendziora:prl2005a}) for all modes.
	
	Now for the matrix elements.  That is, how do the x-rays couple to the modes?  We assume unit coupling to the Higgs and soft modes, the Higgs mode below $T_{s1}$ being the analog of the soft mode above $T_{s1}$.  But for the Goldstone mode, which only exists below $T_{s1}$, we set the coupling constant to $Q^2$ \cite{Fleury}.  So, above $T_{s1}$ we have
	\begin{equation}
	T[1-\tilde{\omega}_s\tan^{-1}{\frac{1}{\tilde{\omega}_s}}]
	\end{equation}
	and below $T_{s1}$ we have
	\begin{equation}
	T[1-\tilde{\omega}_H\tan^{-1}{\frac{1}{\tilde{\omega}_H}}+Q^2(1-\tilde{\omega}_G\tan^{-1}{\frac{1}{\tilde{\omega}_G}})]
	\end{equation}
	
	\begin{figure}
		\includegraphics[width=0.75\columnwidth]{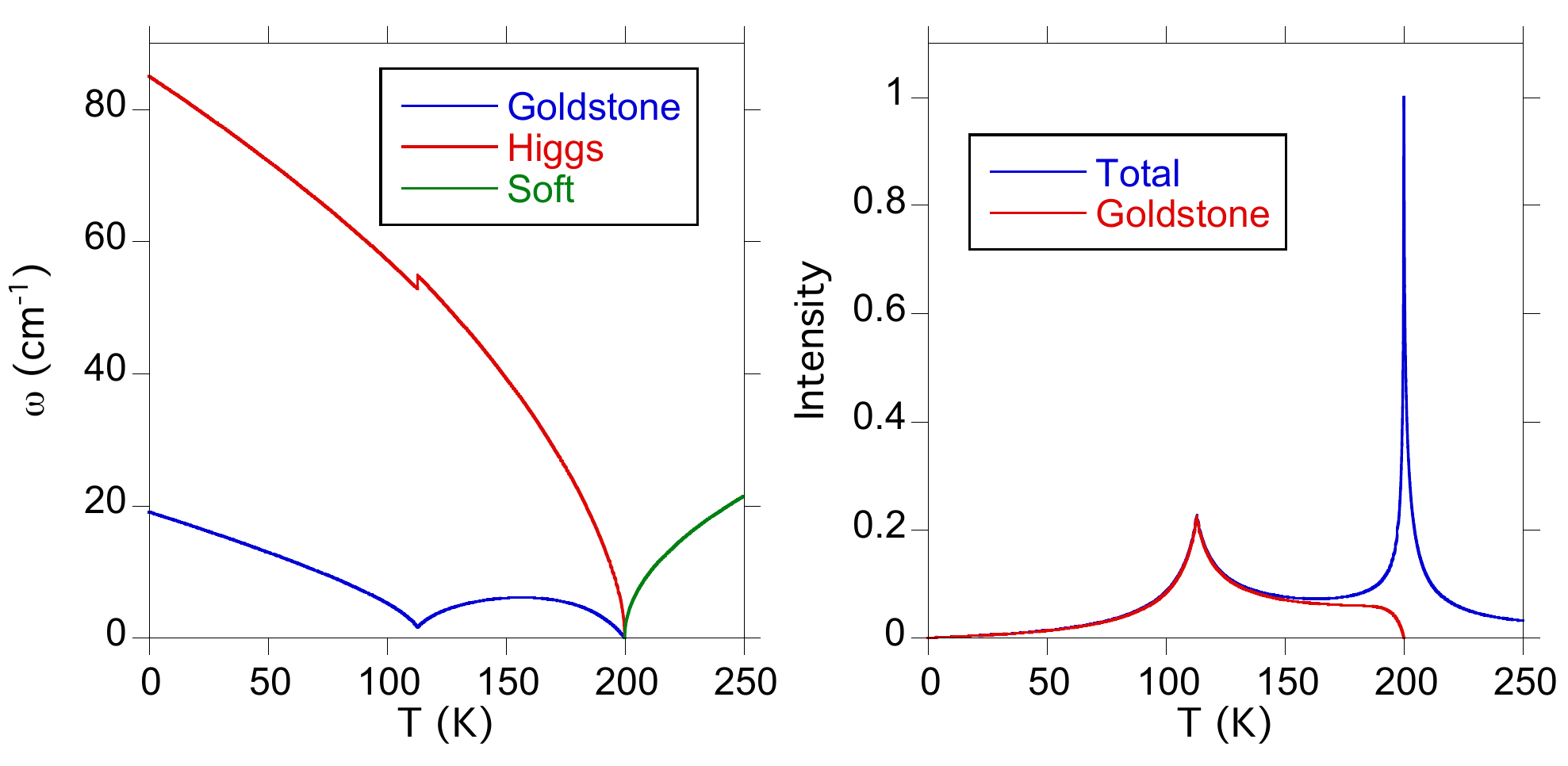}
		\caption{(a) Landau mode energies as a function of $T$ for Cd$_2$Re$_2$O$_7$.  Note the first order jump in the Higgs energy and the dip in the Goldstone energy at $T_{s2}$ (113 K). (b) Landau mode intensities as a function of $T$.  Outside of the critical region near $T_{s1}$ (200 K), the intensity is dominated by the Goldstone intensity.  Note the resemblance of the calculated intensity to the XRD diffuse scattering intensity presented in this paper (Fig.~4(c)).}
		\label{fig-Landau}
	\end{figure}
	
	To evaluate, we choose parameters as in Fig.~3b of \cite{Ivan:2003dl}, with $b_1$=0.3.  We then do the following normalizations.  $a_1$ is some constant times $T-T_{s1}$.  This constant is adjusted so that $T_{s1}$ is 200 K and $T_{s2}$ is 113 K.  Then the mode energies are normalized as stated above (so that the Higgs mode energy is equal to 85 cm$^{-1}$ at $T$=0 as observed by Raman \cite{kendziora:prl2005a}).  Finally, the intensities are normalized by $T_{s1}$.  In Fig.~\ref{fig-Landau}, the resulting mode energies and intensities are shown.  Note the small jump in the Higgs energy and the dip in the Goldstone energy at $T_{s2}$.  Also that the Goldstone intensity completely dominates outside the critical region associated with $T_{s1}$.  As an aside, the Raman data cut-off at about 6 cm$^{-1}$ as noted above.  The prediction is that the Goldstone mode energy should rise above this value at low $T$.  We suggest then that the Raman mode seen at 30 cm$^{-1}$ below $T_{s2}$ could be the Goldstone mode.  This in turn implies that the central peak in the intensity from Raman has more contributions to it than the Goldstone one, and this would presumably be due to elastic scattering from impurities and static short-range structural disorder. 
	
	Finally, some caveats.  First, the behavior well below $T_{s2}$ cannot be taken too seriously since Landau theory is not valid at low $T$ where $Q(T)$ flattens as a function of $T$ (as observed for the Higgs mode by Raman).  Nor for the intensities where the $T/\omega$ approximation for $n(\omega)$ is not valid.  Second, the theory is for a pure $E_u$ model.  In reality, the secondary mode $A_{2u}$ (corresponding to distortions along the $<111>$ trigonal axis orthogonal to $E_u$ distortions) will play some role, and its coupling to $E_u$ is also an anisotropy term in the Landau energy (it does not exist for $I4_122$) \cite{Ivan:2003dl}.   Finally, the critical exponent near $T_{s1}$ is the mean field one.  In reality, experiment finds $\beta$=1/4, not 1/2.  Despite these caveats, Fig.~\ref{fig-Landau} is remarkably similar to the Raman data, and the XRD data reported in this paper.  This brings into question the interpretation of the pump-probe measurements in Ref.~\cite{harter:prl2018a} which claims that a structural soft mode does not exist for Cd$_2$Re$_2$O$_7$.

	\bibliography{X-ray-ML_sources,sm}